\documentclass[preprint,11pt,authoryear]{elsarticle}
\usepackage[authoryear]{natbib}
\usepackage{nomencl}
\makenomenclature
\usepackage{amssymb}
\usepackage{amsmath,amsfonts}
\usepackage{algorithmic}
\usepackage{algorithm}
\usepackage{array}
\usepackage[caption=false,font=normalsize,labelfont=sf,textfont=sf]{subfig}
\usepackage{textcomp}
\usepackage{stfloats}
\usepackage{url}
\usepackage{verbatim}
\usepackage{graphicx}
\usepackage{ragged2e}
\usepackage{etoolbox}
\usepackage{bbm}
\usepackage{mathtools}
\usepackage{placeins}
\usepackage{multirow}
\usepackage{amsthm} 
\newtheorem{theorem}{Theorem}
\usepackage{xcolor}
\usepackage{hyperref}
\usepackage{setspace}
\usepackage[a4paper, left=1.8cm, right=1.8cm, top=2.0cm, bottom=2.4cm]{geometry}
\usepackage{etoolbox} 

\renewcommand\nomgroup[1]{%
  \ifstrequal{#1}{A}{\item[\bfseries Sets and indices:]}{%
  \ifstrequal{#1}{B}{\item[\bfseries Constants:]}{%
  \ifstrequal{#1}{C}{\item[\bfseries Variables:]}{%
  \ifstrequal{#1}{D}{\item[\bfseries Lagrange multipliers:]}{%
  \ifstrequal{#1}{E}{\item[\bfseries Functions:]}{%
  \ifstrequal{#1}{F}{\item[\bfseries Others:]}{}}}}}}}

\makenomenclature
\nomenclature[A,01]{$b$}{Bus index.}
\nomenclature[A]{$\mathcal{I}$}{Set of indices $i$ of generators.}
\nomenclature[A]{$\mathcal{J}$}{Set of indices $j$ of loads.}
\nomenclature[A]{$\mathcal{L}$}{Set of indices $l$ of lines.}
\nomenclature[A]{$\mathcal{K}$}{Set of indices $k$ of credible contingencies.}
\nomenclature[A]{$\mathcal{K}^{OFF}_{i}$}{Set of indices $k$ of the credible contingencies involving the loss of generator $i$.}
\nomenclature[A]{$\mathcal{X}$}{Generator feasibility set.}
\nomenclature[A]{$\mathcal{X}_i$}{Subset of $\mathcal{X}$ related to generator $i$.}
\nomenclature[A]{$\mathcal{Y}$}{Consumer feasibility set.}
\nomenclature[A]{$\mathcal{Y}_j$}{Subset of $\mathcal{Y}$ related to consumer $j$.}

\nomenclature[B, 01]{$\boldsymbol{A}$}{Incidence matrix in the pre-contingency state.}
\nomenclature[B, 02]{$\boldsymbol{A_k}$}{Incidence matrix under contingency $k$.}
\nomenclature[B, 03]{$\boldsymbol{a_k^g}$}{Vector of generator availability statuses under contingency $k$, with each element denoted by $a_{ik}^g$.}
\nomenclature[B, 04]{$\boldsymbol{c}$}{Vector of cost rates offered by generators to provide energy in the pre-contingency state, with each element denoted by $c_i$.}
\nomenclature[B, 05]{$d$}{Inelastic demand.}
\nomenclature[B, 06]{$\boldsymbol{\overline{D}}$}{Vector of upper bounds on the power consumed, with each element denoted by $\overline{D}_j$.}
\nomenclature[B, 10]{$\boldsymbol{F}$}{Vector of line power flow capacities, with each element denoted by $F_l$.}
\nomenclature[B, 11]{$\boldsymbol{\overline{G}}$}{Vector of generator capacities, with each element denoted by $\overline{G}_i$.}
\nomenclature[B, 12]{$\boldsymbol{H}$}{Matrix relating power flows to nodal phase angles in the pre-contingency state.}
\nomenclature[B,13]{$\boldsymbol{q^{d,dn}}$}{Vector of cost rates offered by consumers to provide down-spinning reserve, with each element denoted by $q^{d,dn}_j$.}
\nomenclature[B, 14]{$\boldsymbol{H_k}$}{Matrix relating power flows to nodal phase angles under contingency $k$.}
\nomenclature[B, 15]{$\boldsymbol{M^d}$}{Demand-bus mapping matrix.}
\nomenclature[B, 16]{$\boldsymbol{M^g}$}{Generator-bus mapping matrix.}
\nomenclature[B,17]{$\boldsymbol{q^{d,up}}$}{Vector of cost rates offered by consumers to provide up-spinning reserve, with each element denoted by $q^{d,up}_j$.}
\nomenclature[B,18]{$\boldsymbol{q^{g,dn}}$}{Vector of cost rates offered by generators to provide down-spinning reserve, with each element denoted by $q^{g,dn}_i$.}
\nomenclature[B,19]{$\boldsymbol{q^{g,up}}$}{Vector of cost rates offered by generators to provide up-spinning reserve, with each element denoted by $q^{g,up}_i$.}
\nomenclature[B,21]{$\boldsymbol{\overline{R}^{d,up}}$}{Vector of upper bounds for consumer up-spinning reserves, with each element denoted by $\overline{R}_j^{d,up}$.}
\nomenclature[B,20]{$\boldsymbol{\overline{R}^{d,dn}}$}{Vector of upper bounds for consumer down-spinning reserves, with each element denoted by $\overline{R}_j^{d,dn}$.}
\nomenclature[B,23]{$\boldsymbol{\overline{R}^{g,up}}$}{Vector of upper bounds for generator up-spinning reserves, with each element denoted by $\overline{R}_i^{g,up}$.}
\nomenclature[B,22]{$\boldsymbol{\overline{R}^{g,dn}}$}{Vector of upper bounds for generator down-spinning reserves, with each element denoted by $\overline{R}_i^{g,dn}$.}
\nomenclature[B,24]{$\boldsymbol{w}$}{Vector of rates bid by consumers to buy energy, with each element denoted by $w_j$.}

\nomenclature[C,01]{$\boldsymbol{\theta_0}$}{Vector of pre-contingency nodal phase angles.}
\nomenclature[C,02]{$\boldsymbol{\theta_k}$}{Vector of nodal phase angles under contingency $k$.}
\nomenclature[C,03]{$\boldsymbol{d_0}$}{Vector of pre-contingency nodal consumption levels, with each element denoted by $d_{b0}$.}
\nomenclature[C,04]{$\boldsymbol{d_k}$}{Vector of nodal consumption levels under contingency $k$, with each element denoted by $d_{bk}$.}
\nomenclature[C,05]{$\boldsymbol{g_0}$}{Vector of generator power outputs in the pre-contingency state, with each element denoted by $g_{i0}$.}
\nomenclature[C,06]{$\boldsymbol{g_k}$}{Vector of generator power outputs under contingency $k$, with each element denoted by $g_{ik}$.}
\nomenclature[C,07]{$\boldsymbol{r^{d,dn}}$}{Vector of down-spinning reserves provided by consumers, with each element denoted by $r^{d,dn}_j$.}
\nomenclature[C,08]{$\boldsymbol{r^{d,up}}$}{Vector of up-spinning reserves provided by consumers, with each element denoted by $r^{d,up}_j$.}
\nomenclature[C,09]{$\boldsymbol{r^{g,dn}}$}{Vector of down-spinning reserves provided by generators, with each element denoted by $r^{g,dn}_i$.}
\nomenclature[C,10]{$\boldsymbol{r^{g,up}}$}{Vector of up-spinning reserves provided by generators, with each element denoted by $r^{g,up}_i$.}
\nomenclature[C,11]{$\boldsymbol{x}$}{Vector of optimization variables related to generators.}
\nomenclature[C,12]{$\boldsymbol{x_i}$}{Vector of optimization variables related to generator $i$.}
\nomenclature[C,13]{$\boldsymbol{y}$}{Vector of optimization variables related to consumers.}
\nomenclature[C,14]{$\boldsymbol{y_j}$}{Vector of optimization variables related to consumer $j$.}

\nomenclature[D, 01]{$\boldsymbol{\Pi}$}{Vector of all Lagrange multipliers.}
\nomenclature[D, 02]{$\boldsymbol{\Pi_b}$}{Subvector of $\boldsymbol{\Pi}$ related to bus $b$.}
\nomenclature[D, 03]{$\boldsymbol{\pi_0}$}{Vector of Lagrange multipliers associated with the pre-contingency nodal power balance constraints, with each element denoted by $\pi_{b0}$.}
\nomenclature[D, 04]{$\pi_0$}{Lagrange multiplier associated with the pre-contingency power balance constraint in the single-bus model.}
\nomenclature[D, 05]{$\boldsymbol{\pi_k}$}{Vector of Lagrange multipliers associated with the nodal power balance constraints under contingency $k$, with each element denoted by $\pi_{bk}$.}
\nomenclature[D, 06]{$\pi_k$}{Lagrange multiplier associated with the nodal power balance constraint under contingency $k$ in the single-bus model.}
\nomenclature[D,07]{$\boldsymbol{\pi^{f+}_0}$}{Vector of Lagrange multipliers associated with the lower bounds for the pre-contingency line flows, with each element denoted by $\pi_{l0}^{f+}$.}
\nomenclature[D,08]{$\boldsymbol{\pi^{f-}_0}$}{Vector of Lagrange multipliers associated with the upper bounds for the pre-contingency line flows, with each element denoted by $\pi_{l0}^{f-}$.}

\nomenclature[D,09]{$\boldsymbol{\pi^{f+}_k}$}{Vector of Lagrange multipliers associated with the lower bounds for the line flows under contingency $k$, with each element denoted by $\pi_{lk}^{f+}$.}
\nomenclature[D,10]{$\boldsymbol{\pi^{f-}_k}$}{Vector of Lagrange multipliers associated with the upper bounds for the line flows under contingency $k$, with each element denoted by $\pi_{lk}^{f-}$.}

\nomenclature[E]{$\phi(\cdot)$}{Lagrangian dual function.}
\nomenclature[E]{$\phi_\theta(\cdot)$}{Term of the Lagrangian dual function associated with  $\boldsymbol{\theta_0}$ and $\boldsymbol{\theta_k}$.}
\nomenclature[E]{$\psi^g_{ik}(\cdot)$}{Generator $i$'s revenue fraction under contingency $k$.}
\nomenclature[E]{$\psi^d_{jk}(\cdot)$}{Consumer $j$'s payment fraction under contingency $k$.}

\nomenclature[F,01]{$\pi_{bk}^+$}{Contribution to $p_b^{up}$ due to contingency $k$.}
\nomenclature[F,02]{$\pi_{bk}^-$}{Contribution to $p_b^{dn}$ due to contingency $k$.}
\nomenclature[F,03]{$\boldsymbol{\pi^{f}_0}$}{Vector resulting from summing $\boldsymbol{\pi^{f+}_0}$ and $\boldsymbol{\pi^{f-}_0}$, with each element denoted by $\pi_{l0}^f$.}
\nomenclature[F,03]{$\boldsymbol{\pi^{f}_k}$}{Vector resulting from summing $\boldsymbol{\pi_k^{f+}}$ and $\boldsymbol{\pi_k^{f-}}$, with each element denoted by $\pi^f_{lk}$.}

\nomenclature[F,03]{$b{(i)}$}{Bus of generator $i$.}
\nomenclature[F,04]{$b{(j)}$}{Bus of consumer $j$.}
\nomenclature[F,05]{$C^{d,dn}_j$}{Down-spinning reserve offer cost of consumer $j$.}
\nomenclature[F,06]{$C^{d,t}_j$}{Total cost of consumer $j$.}
\nomenclature[F,07]{$C^{d,up}_j$}{Up-spinning reserve offer cost of consumer $j$.}
\nomenclature[F,08]{$C^{g,dn}_i$}{Down-spinning reserve offer cost of generator $i$.}
\nomenclature[F,09]{$C^{g,e}_i$}{Energy cost of generator $i$.}
\nomenclature[F,10]{$C^{g,s}_i$}{Security charge of generator $i$.}
\nomenclature[F,11]{$C^{g,t}_i$}{Total cost of generator $i$.}
\nomenclature[F,12]{$C^{g,up}_i$}{Up-spinning reserve offer cost of generator $i$.}

\nomenclature[F,13]{$CP$}{Consumer payment.}
\nomenclature[F,14]{$CP_j$}{Payment of consumer $j$.}
\nomenclature[F,15]{$CP^e_j$}{Energy payment of consumer $j$.}
\nomenclature[F,16]{$p^{dn}_b$}{Down-spinning reserve price at bus $b$.}
\nomenclature[F,17]{$p^e$}{Energy price.}
\nomenclature[F,18]{$p_b^e$}{Energy price at bus $b$.}
\nomenclature[F,19]{$p^{f}_l$}{Transmission price for line $l$.}
\nomenclature[F,20]{$p^s$}{Security price.}
\nomenclature[F,21]{$p_b^s$}{Security price at bus $b$.}
\nomenclature[F,22]{$p^{up}$}{Up-spinning reserve price.}
\nomenclature[F,23]{$p^{up}_b$}{Up-spinning reserve price at bus $b$.}
\nomenclature[F,24]{$Profit^d_j$}{Profit of consumer $j$.}
\nomenclature[F,25]{$Profit^g_i$}{Profit of generator $i$.}
\nomenclature[F,26]{$R^{d,dn}_{j}$}{Down-spinning reserve revenue of consumer $j$.}
\nomenclature[F,27]{$R^{d,up}_{j}$}{Up-spinning reserve revenue of consumer $j$.}
\nomenclature[F,28]{$R^{g,dn}_{i}$}{Down-spinning reserve revenue of generator $i$.}
\nomenclature[F,29]{$R^{g,e}_i$}{Energy revenue of generator $i$.}
\nomenclature[F,30]{$R^{g,t}_i$}{Total revenue of generator $i$.}
\nomenclature[F,31]{$R^{g,up}_{i}$}{Up-spinning reserve revenue of generator $i$.}

\nomenclature[F,33]{$U^d_j$}{Utility of consumer $j$.}

\journal{EJOR}

\begin{document}

\renewcommand{\nompreamble}{This section lists the main notation used throughout the paper. Note that superscript “*” stands for optimal value.}

\begin{frontmatter}

\title{A Causation-Based Framework for Pricing and Cost Allocation of Energy, Reserves, and Transmission in Modern Power Systems}

\author[1]{Luíza  Ribeiro}
\author[1]{Alexandre Street\corref{cor1}}
\cortext[cor1]{Corresponding author}
\ead{street@puc-rio.br}
\author[2]{José M. Arroyo}
\author[3]{Rodrigo Moreno}

\affiliation[1]{organization={Electrical Engineering Department, Pontifical Catholic University of Rio de Janeiro},
                city={Rio de Janeiro},
                country={Brazil}}

\affiliation[2]{organization={E.T.S.I. Industrial, Universidad de Castilla-La Mancha},
                city={Ciudad Real},
                country={Spain}}
                
\affiliation[3]{organization={
FCFM, University of Chile},
                city={Santiago},
                country={Chile}}


\begin{abstract}

The increasing vulnerability of power systems has heightened the need for operating reserves to manage contingencies such as generator outages, line failures, and sudden load variations. Unlike energy costs, driven by consumer demand, operating reserve costs arise from addressing the most critical credible contingencies—prompting the question: how should these costs be allocated through efficient pricing mechanisms? As an alternative to previously reported schemes, this paper presents a new causation-based pricing framework for electricity markets based on contingency-constrained energy and reserve scheduling \mbox{models}. Major salient features include a novel security charge mechanism along with the explicit definition of prices for up-spinning reserves, down-spinning reserves, and transmission services. These features ensure more comprehensive and efficient cost-reflective market operations. Moreover, the proposed nodal pricing scheme yields revenue adequacy and neutrality while promoting reliability incentives for generators based on the cost-causation principle. An additional salient aspect of the proposed framework is the economic incentive for transmission assets, which are remunerated based on their use to deliver energy and reserves across all contingency states. Numerical results from two case studies illustrate the performance of the proposed pricing scheme.
\end{abstract}


\end{frontmatter}

\newpage

\section{Introduction}

Ensuring system security—defined as the ability to withstand credible contingencies such as generator outages, line failures, or sudden load variations without involuntary load shedding—has long been a fundamental priority for the power industry \citep{Arroyo_2005}. Therefore, in modern restructured power systems, besides energy, ancillary services have become indispensable for meeting diverse security requirements. Among these, operating reserves (ORs) are crucial for maintaining system reliability by balancing supply and demand. However, due to the more frequent natural hazards and the massive penetration of low-inertia generation, power systems worldwide are experiencing a growing vulnerability. As a result, ORs represent an increasing share of operational costs \citep{Shi2023, BADESA2024, BADESA2025}.

Despite the increasing reliance on ancillary services, existing electricity markets have failed to establish adequate incentives or frameworks to meet the evolving challenges of modern power systems \citep{Billimoria2020}. Markets generally follow either joint or sequential designs for energy and reserve clearing \citep{Galiana_2005, RIBEIRO2023}. Joint markets, like those run by PJM and CAISO in the United States, co-optimize energy and reserves simultaneously. Conversely, sequential markets, common in Europe, clear energy and reserves separately. While sequential markets aim to optimize reserves at a portfolio level, they often rely on \emph{ad hoc} allocation of transmission capacity for reserves, leading to inefficiencies and suboptimal outcomes \citep{ACER_CACM}. In contrast, joint markets typically achieve higher social welfare by better capturing the interdependencies between these products, as evidenced by several studies \citep{Arroyo_2005, Galiana_2005, Aganagic1998, Deqiang2003, Tong2004, Bouffard2005, wong2007, Wang2009,  Karangelos2012, Morales2012, Book_Fundamentals_2004}. However, many joint markets rely on static, exogenously defined reserve requirements, such as PJM’s largest-generator rule \citep{PJM_Manual_11} or CAISO’s proportional load-based allocation \citep{CAISO_Market_Report_2019}. These static methods fail to reflect the dynamic nature of system operations, significantly influencing market-clearing outcomes and reserve pricing. Moreover, reserve deliverability during contingencies is often overlooked, leading to \emph{ad hoc} solutions like zonal divisions with separate reserve requirements and prices \citep{Shi2023}.

Another persistent challenge lies in the cost allocation of reserves. Most Independent System Operators allocate reserve costs proportionally among load-serving entities based on energy consumption, which raises concerns about fairness and efficiency \citep{Shi2023, BADESA2024, BADESA2025}. Proportional allocation often results in cross-subsidies, as it fails to link costs directly to the entities responsible for reserve requirements. As far as the authors are aware, among existing markets, cost-causation principles for reserve cost allocation are solely applied by the Australian Energy Market Operator (AEMO) across its two independent systems: the Wholesale Electricity Market (WEM) and the National Electricity Market (NEM). However, while the WEM employs a sequential approach that better aligns costs with responsible entities, the NEM still relies on proportional allocation \citep{AEMO2023a}, perpetuating inefficiencies.

In light of these challenges, numerous studies have proposed deterministic frameworks, such as those relying on the N-1 security standard, as well as probabilistic approaches to improve reserve pricing and scheduling  \citep{Arroyo_2005, Galiana_2005, Bouffard2005, wong2007, Wang2009, Karangelos2012, Morales2012, ONEILL2005269,  MAYS2021712, BADESA_MATAMALA2023, BYERS2023351, nuran2023, AHUNBAY2024605, Street10970215}. These approaches derive energy and reserve prices from the dual variables of power balance constraints, ensuring that prices reflect the marginal costs of maintaining reliability. However, most frameworks assume ancillary services costs are fully borne by consumers. Additionally, existing models fail to incorporate transmission revenues into reserve pricing.

Early discussions on reserve cost allocation focused on generator outages. Proportional cost allocation based on generating units' capacity and unavailability was proposed in \citet{Goran2000}. In \citet{KIRBY2003}, this idea was expanded with mechanisms incorporating historical outages and dispatch size. However, using historical outages as a proxy for predicting future events is suboptimal due to their low probability of occurrence, leading to erratic charges and misalignment between prospective reserve procurement and retrospective cost allocation. Specifically, if no outages occur, reserve costs lack clear entities to charge \citep{BADESA_MATAMALA2023}. More recent studies, such as \citet{Xiang2023} and \citet{lIU2023}, emphasize renewables and consumption variabilities as factors influencing reserve costs, relying on historical data. However, these variabilities are not the dominant driver of reserve needs in low-inertia systems \citep{BADESA2024}, and historical data often fail to predict future contingencies, underscoring the need for adaptive, causation-based allocation methods. 

In \citet{BADESA2024}, the costs of ORs in low-inertia systems are shown to be primarily driven by large generator outages. The study evaluates proportional, Shapley value, and nucleolus methods, identifying the nucleolus as the most fair due to its ability to avoid cross-subsidies and incentivize cooperative behavior. Building on this finding, cost-causation principles are advocated in \citet{BADESA2025} to ensure reserve costs are borne by those creating the need for ancillary services. Furthermore, in \citet{BADESA2025}, the authors highlight the need for cost-reflective frameworks that incentivize responsible behavior and investment in future grids. However, to date, no study combines a unified pricing model for energy, reserves, and transmission with a causation-based cost allocation methodology, leaving an important gap in the literature.

This paper addresses these issues by introducing a unified framework for pricing and cost allocation of energy, reserves, and transmission services, grounded in causation-based principles. More specifically, the objective of this work is to extend the findings of \citet{Arroyo_2005}—where uniform prices for energy and reserves were derived for the case of contingency-constrained models—to incorporate the cost-causation principle. The proposed approach gives rise to a pricing system that aligns consumers' payments with generation and transmission revenues while ensuring cost recovery for market participants. Thus, we contribute to knowledge with the following new concepts: 

\begin{enumerate}
    \item A security charge that complements the price of energy and reserves proposed in \citet{Arroyo_2005} to efficiently allocate reserve costs directly to the entities responsible for causing them,
    \item the differentiation between up- and down-spinning reserve prices based on the system's opportunity cost, and
    \item a pricing system that properly remunerates transmission lines for the spare capacity used to ensure reserve deliverability through the network.  
\end{enumerate}

These newly proposed concepts rely on Lagrange multipliers, ensuring a transparent, fair, and efficient cost allocation. Based on those concepts, we achieve a consistent settlement process that adheres to key axiomatic principles for energy and reserve contingency-constrained models. These principles include: 1) revenue adequacy, ensuring non-negative profit (surplus) for producers and consumers, and 2) revenue neutrality, whereby total consumers' payment equals total generation and transmission revenue.

It is important to note that while this study focuses on generation-driven cost causation, the same principles apply to demand-driven cost causation. Although frequency drops are the primary concern in most power systems, over-frequency events can be modeled analogously by considering sudden load disconnections instead of generation outages. 

The remainder of this paper is structured as follows. Section \ref{sect:First_phase_1} develops the pricing framework for a simplified model without network constraints, considering only generator contingencies to place the focus on the concept of security charges and highlight their role in efficient market operations. In Section \ref{sect:up and down reserves}, the model is extended to incorporate network constraints, demand response, and both generator and line outages. This extension reinforces the necessity of security charges in more complex systems, examines their impact on the revenues of different market participants, including transmission assets, and introduces a novel approach for separately pricing up- and down-spinning reserves. Each section presents the corresponding optimization model for market clearing, introduces an example that is addressed using previous results from \citet{Arroyo_2005}, develops the proposed pricing framework, and provides and discusses the new results using the same example previously considered in the section. Section \ref{sect:conclusion} draws relevant conclusions and suggests directions for future research. Finally, the nomenclature is given in Appendix A, and mathematical proofs are provided in Appendices B--E.

\section{Introducing the Notion of Security Charges}
\label{sect:First_phase_1}

This section introduces the concept of security charges under marginal pricing through a simplified contingency-constrained market-clearing model for energy and reserves. Briefly, the security charge concept is a cost allocation mechanism that reflects the different contributions of generators to the endogenously defined reserve needs, thereby addressing the limitations of existing uniform pricing in contingency-constrained models. Larger scheduled generators, whose potential outages create higher reserve needs and costs, incur higher charges, leading to a pricing framework that more accurately aligns costs with the system's opportunity costs. This concept will be further developed in the following sections.

To that end, we first formulate the stylized optimization model used to introduce the security charge concept. Then, we revisit the existing uniform pricing framework in \citet{Arroyo_2005} and illustrate its limitations. Finally, we develop the notion of security charges based on the study of the Lagrangian dual function and how each term in this function should be attributed to each agent to ensure the cost-causation principle. The benefits of the proposed pricing system are illustrated through the assessment with the method described in \citet{Arroyo_2005}.

\subsection{Optimization Model}
For expository purposes, generation outages are considered using a practical deterministic security criterion \citep{wood_2013}, whereas a single-bus, single-period model is employed, considering an inelastic demand and linear offer cost functions, with a focus on spinning or synchronized reserves. In this simplified framework, only one type of reserve, namely upward spinning reserves, is required to ensure system security. Thus, the market-clearing procedure is given by the following contingency-constrained model, which is an instance of linear programming:

\begin{align}
& \min\limits_{\boldsymbol{x} \geq \boldsymbol{0}} 
 \quad  \boldsymbol{c}^{\top} \boldsymbol{g_0} + (\boldsymbol{q^{g,up}})^{\top} \boldsymbol{r^{g,up}} \label{eq:SMobj} \\  
& \text{subject to:} &  \nonumber \\
& \mathbbm{1}^{\top} \boldsymbol{g_0} = d \quad (\pi_0) \label{eq:SMcon2} \\
&\mathbbm{1}^{\top} \boldsymbol{g_k} = d \quad (\pi_k), \quad 
 \quad  \quad \quad  \ \ \forall k \in \mathcal{K} \label{eq:SMcon3} \\
& \boldsymbol{g_k} \leq \mathbb{D}(\boldsymbol{a^g_k})(\boldsymbol{g_0 + r^{g,up}}), \quad \forall k \in \mathcal{K} \label{eq:SMcon4} \\
& \boldsymbol{g_0} + \boldsymbol{r^{g,up}} \leq \boldsymbol{\overline{G}} \label{eq:SMcon5}\\
&\boldsymbol{r^{g,up}}  \leq \boldsymbol{\overline{R}^{g,up}} \label{eq:SMcon6} 
\end{align}

\noindent
where $\boldsymbol{x} = \{\boldsymbol{g_0}, \boldsymbol{r^{g,up}}, \left\{ \boldsymbol{g_k} \right\}_{k \in \mathcal{K}} \}$, $\mathbb{D}(\cdot)$ denotes the diagonal matrix operator, and $\mathbbm{1}$ represents the all-ones vector.

The objective function to be minimized \eqref{eq:SMobj} consists of the sum of the costs for generating power and providing up-spinning reserves offered by the generators.

Constraints \eqref{eq:SMcon2} and \eqref{eq:SMcon3} ensure the power balance between generation and consumption under both pre-contingency and contingency states, respectively. Accordingly, the total output of all generators under every state must equal the load demand, which is assumed to be constant. Note that $\pi_0$ and $\pi_k$ represent the associated Lagrange multipliers.

Constraint \eqref{eq:SMcon4} relates the up-spinning reserve contributions to the power levels produced under the pre-contingency and contingency states. Lastly, constraints \eqref{eq:SMcon5} and \eqref{eq:SMcon6} set the operational limits of the generator outputs and reserves.

\subsection{Challenges of the Existing Pricing Framework}
\label{sect: sm_example_arroyo}
For contingency-constrained models such as problem \eqref{eq:SMobj}--\eqref{eq:SMcon6}, prices under a marginal pricing framework \citep{SCHWEPPE} can be computed using the methodology described in \citet{Arroyo_2005}, which is based on the use of Karush-Kuhn-Tucker (KKT) optimality conditions.

In \citet{Arroyo_2005}, the energy price is defined as the sum of the Lagrange multipliers corresponding to the power balance in both pre-contingency and contingency states. Moreover, according to \citet{Arroyo_2005}, reserves are priced through the so-called security price, which is defined as the sum of the Lagrange multipliers related to the power balance equations under contingency. As a result, for problem \eqref{eq:SMobj}--\eqref{eq:SMcon6}, generators collect revenues for their pre-contingency power outputs and scheduled up-spinning reserves using the aforementioned energy and security prices, respectively. Analogously, consumers are charged the energy price for their consumption. Table \ref{tab:SM_pricing_system_arroyo} summarizes the prices and settlement according to \citet{Arroyo_2005} for the market-clearing model \eqref{eq:SMobj}--\eqref{eq:SMcon6}.

\begin{table}[!ht]
\centering
\captionsetup{justification=centering, width=\linewidth}
\caption{Pricing system and settlement of \citet{Arroyo_2005} for problem \eqref{eq:SMobj}--\eqref{eq:SMcon6}}
\label{tab:SM_pricing_system_arroyo}
\resizebox{0.5\textwidth}{!}{ 
\begin{tabular}{|l|l|}
\hline
\multirow{2}{*}{\textbf{Energy Price}} &
  \multirow{2}{*}{$ p^e = \pi^*_0 + \sum\limits_{k \in \mathcal{K}} \pi^*_k$} \\
                                                    &                                             \\ \hline
\multirow{2}{*}{\textbf{Security Price}} &
  \multirow{2}{*}{$p^s = \sum\limits_{k \in \mathcal{K}} \pi^*_k $} \\
                                                    &                                             \\ \hline
\multirow{2}{*}{\textbf{Generation Energy Revenue}} & \multirow{2}{*}{$R^{g,e}_i = p^e g_{i0}$}       \\
                                                    &                                             \\ \hline
\multirow{2}{*}{\textbf{\begin{tabular}[c]{@{}l@{}}Generation Up-Spinning \\ Reserve Revenue\end{tabular}}} &
  \multirow{2}{*}{$R^{g,up}_i = p^s r^{g,up}_{i}$} \\
                                                    &                                             \\ \hline
\multirow{2}{*}{\textbf{Generation Total Revenue}} &
  \multirow{2}{*}{$ R^{g,t}_i = R^{g,e}_i + R^{g,up}_i$} \\
                                                    &                                             \\ \hline
\multirow{2}{*}{\textbf{Consumer Payment}}          & \multirow{2}{*}{$CP = p^e d $}              \\
                                                    &                                             \\ \hline
\multirow{2}{*}{\textbf{Generation Energy Cost}}    & \multirow{2}{*}{$C_i^{g,e} = c_ig_{i0}$}        \\
                                                    &                                             \\ \hline
\multirow{2}{*}{\textbf{\begin{tabular}[c]{@{}l@{}}Generation Up-\\ Spinning Reserve Cost\end{tabular}}} &
  \multirow{2}{*}{$C_i^{g,up} = q_i^{g,up}r_i^{g,up}$} \\
                                                    &                                             \\ \hline
\multirow{2}{*}{\textbf{Generation Total Cost}}     & \multirow{2}{*}{$C^{g,t}_i = C^{g,e}_i + C_i^{g,up}$}    \\
                                                    &                                             \\ \hline
\multirow{2}{*}{\textbf{Generation Profit}}         & \multirow{2}{*}{$Profit^g_i = R^{g,t}_i - C^{g,t}_i$} \\
                                                    &                                             \\ \hline
\end{tabular}
}
\end{table}

We now apply this pricing framework to an illustrative example involving three generators. Generation data are provided in Table \ref{tab:gen_data_1bus}.  The load is $120$ MW. Three credible contingencies are considered here, each defined by the outage of each generator. Using the simplex algorithm of CPLEX, problem \eqref{eq:SMobj}--\eqref{eq:SMcon6} has been solved to optimality. Table \ref{tab:schedules_1bus} presents the optimal solution, which features a total cost equal to $\$5,800$. As can be seen, the optimization prioritizes energy and reserve offers in reverse order, scheduling the least-cost generator (generator 1) exclusively to supply power in the pre-contingency state. In contrast, generators 2 and 3 are used for both power provision in the pre-contingency state and up-spinning reserve. 
Note that the up-spinning reserve contributions of generators 2 and 3, which are respectively equal to their corresponding upper bounds, amount to the pre-contingency power output of generator 1. It is also worth mentioning that the up-spinning reserve contributions of generators 2 and 3 exceed the reserve needed to guard against the loss of generators 3 and 2, respectively. In other words, the outage of generator 1 is the critical contingency state.

\begin{table}[h!]
\centering
\caption{Single-bus example -- Generation data}
\label{tab:gen_data_1bus}
\renewcommand{\arraystretch}{1.2} 
\resizebox{0.4 \textwidth}{!}{
\begin{tabular}{ccccc}
\hline
\textbf{$i$} &
  \begin{tabular}[c]{@{}c@{}}$\overline{G}_i$\\ (MW)\end{tabular} &
  \begin{tabular}[c]{@{}c@{}}$\overline{R}_i^{g,up}$\\ (MW)\end{tabular} &
  \begin{tabular}[c]{@{}c@{}}$c_i$\\ (\$/MWh)\end{tabular} &
  \begin{tabular}[c]{@{}c@{}}$q_i^{g,up}$\\ (\$/MW)\end{tabular} \\ \hline
\textbf{1} & 100                    & 50 & \textcolor{white}{0}20  & \textcolor{white}{0}2  \\
\textbf{2} & \textcolor{white}{0}60 & 30 & \textcolor{white}{0}50  & \textcolor{white}{0}5  \\
\textbf{3} & \textcolor{white}{0}70                     & 35 & 100 & 10 \\ \hline
\end{tabular}
}
\end{table}

\begin{table}[H]
\caption{Single-bus example -- Optimal results (MW)}
\label{tab:schedules_1bus}
\centering
\resizebox{0.7\textwidth}{!}{
\begin{tabular}{cccccc}
\hline
\multicolumn{3}{c}{}                                            & \multicolumn{3}{c}{\textbf{Generation under Contingency}}                \\ \cline{4-6} 
\textbf{Generator} &
  \textbf{\begin{tabular}[c]{@{}c@{}}Pre-Contingency \\ Generation\end{tabular}} &
  \textbf{\begin{tabular}[c]{@{}c@{}}Up-Spinning\\ Reserve\end{tabular}} &
  \textbf{\begin{tabular}[c]{@{}c@{}}Outage of\\ Generator 1\end{tabular}} &
  \textbf{\begin{tabular}[c]{@{}c@{}}Outage of\\ Generator 2\end{tabular}} &
  \textbf{\begin{tabular}[c]{@{}c@{}}Outage of\\ Generator 3\end{tabular}} \\ \hline
\textbf{1}     & \textcolor{white}{0}65 & \textcolor{white}{0}0 & \textcolor{white}{00}0 & \textcolor{white}{0}65 & \textcolor{white}{0}65 \\
\textbf{2}     & \textcolor{white}{0}30 & 30                    & \textcolor{white}{0}60 & \textcolor{white}{00}0 & \textcolor{white}{0}55 \\
\textbf{3}     & \textcolor{white}{0}25 & 35                    & \textcolor{white}{0}60 & \textcolor{white}{0}55 & \textcolor{white}{00}0 \\ \hline
\textbf{Total} & 120                    & 65                    & 120                    & 120                    & 120                    \\ \hline
\end{tabular}
}
\end{table}

\begin{table}[H]
\caption{ Single-bus example -- Lagrange multipliers (\$/MWh)}
\label{tab:lagrange_1bus}
\centering
\resizebox{0.5 \textwidth}{!}{
\begin{tabular}{cccc}
\hline
\textbf{\begin{tabular}[c]{@{}c@{}}Pre-Contingency\\ State\end{tabular}} &
  \textbf{\begin{tabular}[c]{@{}c@{}}Outage of\\ Generator 1\end{tabular}} &
  \textbf{\begin{tabular}[c]{@{}c@{}}Outage of\\ Generator 2\end{tabular}} &
  \textbf{\begin{tabular}[c]{@{}c@{}}Outage of\\ Generator 3\end{tabular}} \\ \hline
20 &
  80 &
  0 &
  0 \\ \hline
\end{tabular}
}
\end{table}
\FloatBarrier

The Lagrange multipliers associated with the power balance equations in the pre-contingency and contingency states are listed in Table \ref{tab:lagrange_1bus}. For the outages of generators 2 and 3, the associated multipliers are zero, as the loss of these generators does not constrain the system. In contrast, the multiplier for the outage of generator 1 is different from zero as an infinitesimal perturbation of the corresponding power balance equation would yield a different pre-contingency dispatch and reserve schedule, thereby resulting in a change in the value of the objective function.

Tables \ref{tab:price_1bus} and \ref{tab:revenues_1bus_3Gen_arroyo} respectively report the prices and settlement for the generators as per the definitions of Table \ref{tab:SM_pricing_system_arroyo}. Also, according to Table \ref{tab:SM_pricing_system_arroyo}, the consumer payment is given by $p^ed = 100 \times 120 = \$ 12,000$. Interestingly, the pricing scheme proposed in \citet{Arroyo_2005} yields a total generation revenue exceeding the total consumer payment by \$5,200, a missing money generated by the pricing system, as highlighted in Table \ref{tab:system_settl_1bus_JM}. This mismatch is addressed by the novel approach described in the next section.

\begin{table}[h!]
\captionsetup{justification=centering, width=\linewidth}
\caption{Single-bus example -- Prices according to \citet{Arroyo_2005} (\$/MWh)}
\label{tab:price_1bus}
\centering
\resizebox{0.3 \textwidth}{!}{
\begin{tabular}{cc}
\hline
\multicolumn{1}{c}{\textbf{\begin{tabular}[c]{@{}c@{}}Energy Price, \\ $p^e$\end{tabular}}} &
  \multicolumn{1}{c}{\textbf{\begin{tabular}[c]{@{}c@{}}Security Price,\\ $p^s$\end{tabular}}} \\ \hline
100 &
  80 \\ \hline
\end{tabular}
}
\end{table}

\begin{table}[h!]
\captionsetup{justification=centering, width=\linewidth}
\caption{Single-bus example -- Generation settlement according to \citet{Arroyo_2005} (\$)}
\label{tab:revenues_1bus_3Gen_arroyo}
\centering
\resizebox{0.6 \textwidth}{!}{%
\renewcommand{\arraystretch}{1.2} 
\begin{tabular}{cccccccc}
\hline
${i}$          & ${R^{g,e}_i}$             & ${R^{g,up}_i}$ & ${R^{g,t}_i}$             & ${C^{g,e}_i}$  & ${C^{g,up}_i}$ & ${C^{g,t}_i}$  & ${Profit^g_i}$            \\ \hline
\textbf{1} & \textcolor{white}{0}6,500 & \textcolor{white}{000,}0 & \textcolor{white}{0}6,500 & 1,300 & \textcolor{white}{00}0 & 1,300 & \textcolor{white}{0}5,200 \\
\textbf{2}     & \textcolor{white}{0}3,000 & 2,400          & \textcolor{white}{0}5,400 & 1,500          & 150            & 1,650          & \textcolor{white}{0}3,750 \\
\textbf{3}     & \textcolor{white}{0}2,500 & 2,800          & \textcolor{white}{0}5,300 & 2,500          & 350            & 2,850          & \textcolor{white}{0}2,450 \\ \hline
\textbf{Total} & \textbf{12,000}           & \textbf{5,200} & \textbf{17,200}           & \textbf{5,300} & \textbf{500}   & \textbf{5,800} & \textbf{11,400}           \\ \hline
\end{tabular}
}
\end{table}

\begin{table}[h!]
\captionsetup{justification=centering, width=\linewidth}
\caption{Single-bus example -- System settlement according to \citet{Arroyo_2005} (\$)}
\label{tab:system_settl_1bus_JM}
\centering
\resizebox{0.3\textwidth}{!}{ 
\begin{tabular}{cc}
\hline
\textbf{Generation Revenue} & \textcolor{white}{-}17,200 \\
\textbf{Consumer Payment}   & \textcolor{white}{-}12,000 \\ \hline
\textbf{Balance}            & \textcolor{white}{0}-5,200 \\ \hline
\end{tabular}
}
\end{table}

\subsection{Proposed Pricing System}
\label{sect: pricing_SM}
As an alternative to \citet{Arroyo_2005}, the proposed pricing system acknowledges the contribution of each generator to keep the power balance under each contingency state. To that end, we explore the structure of the Lagrangian dual (LD) problem to ensure a revenue-neutral market clearing where the total generation revenue is equal to the consumer payment. The LD problem can be viewed as a price-based coordination approach among market agents that, under specific conditions, yields the optimal solution to the problem under consideration \citep{Kluwer}. Note that, within such a framework, Lagrange multipliers play the role of prices. For problem \eqref{eq:SMobj}--\eqref{eq:SMcon6}, the LD function is built as follows \citep{Kluwer}:

\begin{equation}
    \begin{aligned}
        \label{eq:SM_Lagrangian1}
         \phi(\boldsymbol{\Pi}) = \underset{\boldsymbol{x} \in \mathcal{X}}{\min} \ \ 
        & \boldsymbol{c}^{\top} \boldsymbol{g_0} + \left(\boldsymbol{q^{g,up}}\right)^{\top} \boldsymbol{r^{g,up}}  
        + \pi_0 \left(d - \mathbbm{1}^{\top} \boldsymbol{g_0}\right) \\
        &+ \sum\limits_{k \in \mathcal{K}} \pi_k \left(d - \mathbbm{1}^{\top} \boldsymbol{g_k}\right) 
    \end{aligned}
\end{equation}

\noindent
where $\boldsymbol{\Pi} = \{\pi_0, \{\pi_k\}_{k \in \mathcal{K}}\}$ and $\mathcal{X}$ includes constraints \eqref{eq:SMcon4}--\eqref{eq:SMcon6} and $\boldsymbol{x} \geq \boldsymbol{0}$. 

Taking the terms that do not depend on $\boldsymbol{x}$ out of the minimization problem and factoring out yields: 

\begin{equation}
    \begin{aligned}
        \label{eq:SM_Lagrangian2}
        \phi(\boldsymbol{\Pi})  = &\left(\pi_0 + \sum\limits_{k \in \mathcal{K}} \pi_k\right) d 
        +\underset{\boldsymbol{x} \in \mathcal{X}}{\min} \Bigg\{ \left(\boldsymbol{c}^{\top} - \pi_0 \mathbbm{1}^{\top} \right) \boldsymbol{g_0}\\
        &+\left(\boldsymbol{q^{g,up}}\right)^{\top}\boldsymbol{r^{g,up}}
        -\sum\limits_{k \in \mathcal{K}} \pi_k \mathbbm{1}^{\top} \boldsymbol{g_k} \Bigg\}
    \end{aligned}
\end{equation}

The minimization problem in \eqref{eq:SM_Lagrangian2} is generator-wise separable, as no constraints are coupling the actions of generators. Also, by factoring out the negative sign, we can transform the $min$ operator into a $max$ operator. Thus, rewriting Equation \eqref{eq:SM_Lagrangian2} accordingly, we have Equation \eqref{eq:SM_Lagrangian3} as follows:

\begin{equation}
    \begin{aligned}
        \label{eq:SM_Lagrangian3}
        \phi(\boldsymbol{\Pi}) = &\left(\pi_0 + \sum\limits_{k \in \mathcal{K}} \pi_k\right) d 
         -\sum\limits_{i \in \mathcal{I}}\underset{\boldsymbol{x_i} \in \mathcal{X}_i}{\max}\Bigg[ \left(\pi_{0} - c_i \right) g_{i0} \\
         &- q_i^{g,up}r_i^{g,up} 
         + \sum\limits_{k \in \mathcal{K}}\psi^g_{ik}(\pi_k, \boldsymbol{x_i})\Bigg]
    \end{aligned}
\end{equation}

\noindent
where $\boldsymbol{x_i}= \{g_{i0}, r_i^{g,up}\}$, $ \mathcal{X}_i$ is the subset of constraints related to $\boldsymbol{x_i}$, i.e., constraints \eqref{eq:SMcon5}, \eqref{eq:SMcon6}, and $\boldsymbol{x_i} \geq \boldsymbol{0}$,  whereas $\psi^g_{ik}(\pi_k, \boldsymbol{x_i})$ stands for the revenue fraction of generator $i$ under contingency $k$ due to its best response against post-contingency-state price $\pi_k$, given its pre-contingency dispatch and reserve schedule, $\boldsymbol{x_i}$, and availability status at state $k$. Note that $\psi^g_{ik}(\pi_k, \boldsymbol{x_i})$ can be cast as: 

\vspace{-1.5ex}
\begin{align}
&\psi^g_{ik}(\pi_k, \boldsymbol{x_i}) =  \max\limits_{g_{ik} \geq 0} \quad \pi_kg_{ik} \label{eq:genobj} \\  
& \text{subject to:} &  \nonumber \\
& g_{ik} \leq a^g_{ik}\left(g_{i0}+r_i^{g,up} \right)
\label{eq:bound_gen}
\end{align}

Based on KKT conditions, $\pi_k$ can only take non-negative values at the optimal solution to \eqref{eq:SMobj}--\eqref{eq:SMcon6}. As a consequence, we have two possible outputs for $\psi^{g}_{ik}(\pi_k, \boldsymbol{x_i})$: 1) If $\pi_k$ is positive, $g_{ik}^*$ will take its maximum possible value, given by $g_{ik}^* = a^g_{ik}\left(g_{i0} + r_i^{g,up}\right)$, and $\psi^{g}_{ik}(\pi_k, \boldsymbol{x_i})$ will be equal to $\pi_k a^g_{ik}\left(g_{i0} + r_i^{g,up}\right)$;  2) If $\pi_k$ is zero, $\psi^{g}_{ik}(\pi_k, \boldsymbol{x_i})$ will be $0$. Thus, $\psi^{g}_{ik}(\pi_k, \boldsymbol{x_i})$ will only take values different from zero when $\pi_k$ is positive. Hence, $\psi^{g}_{ik}(\pi_k, \boldsymbol{x_i})$ can be equivalently replaced with $\pi_k a^g_{ik}\left(g_{i0} + r_i^{g,up}\right)$. Rewriting \eqref{eq:SM_Lagrangian3} accordingly, we have Equation \eqref{eq:SM_Lagrangian4} as follows:  

\begin{equation}
    \begin{aligned}
        \label{eq:SM_Lagrangian4}
        \phi(\boldsymbol{\Pi}) = &\left(\pi_0 + \sum\limits_{k \in \mathcal{K}}\pi_k\right) d  
        -\sum\limits_{i \in \mathcal{I}} \max\limits_{\boldsymbol{x_i} \in \mathcal{X}_i}\Bigg[\left(\pi_{0} - c_i \right)g_{i0} \\
        &-q_i^{g,up} r_i^{g,up} 
        + \sum\limits_{k \in \mathcal{K}} \pi_k a^g_{ik}\left(g_{i0} + r_i^{g,up}\right) \Bigg]
    \end{aligned}
\end{equation}

By rearranging Equation \eqref{eq:SM_Lagrangian4}, we obtain Equation \eqref{eq:SM_Lagrangian5}: 

\begin{equation}
    \begin{aligned}
        \label{eq:SM_Lagrangian5}
        \phi(\boldsymbol{\Pi}) = &\left(\pi_0 + \sum\limits_{k \in \mathcal{K}} \pi_k \right) d 
        - \sum\limits_{i \in \mathcal{I}} \max\limits_{\boldsymbol{x_i} \in \mathcal{X}_i} \Bigg[ 
        \left(\pi_{0} + \sum\limits_{k \in \mathcal{K}} \pi_k a^g_{ik} - c_i \right) g_{i0} \\
        &+ \left(\sum\limits_{k \in \mathcal{K}} \pi_k a^g_{ik} - q_i^{g,up} \right) r_i^{g,up} \Bigg]
    \end{aligned}
\end{equation}

The first term in the right-hand side of \eqref{eq:SM_Lagrangian5} represents the consumer payment. The terms in square brackets represent the profit of generator $i$, which is made up of the net revenue, i.e., revenue minus cost, from both selling energy and providing up-spinning reserve. Note that $a^g_{ik}$ is a binary parameter that is equal to $0$ when generator $i$ is out of service under contingency state $k$. Thus, generator $i$ solely collects revenues at $\pi_k$ for contingency states $k$ in which this generator is available. This is equivalent to saying that generator $i$ gets paid $\pi_k$ for every contingency state $k$ while being charged $\pi_k$ in the contingency states in which this generator is out of service. Rewriting \eqref{eq:SM_Lagrangian5} accordingly, we have Equation \eqref{eq:SM_Lagrangian6}:

\begin{equation}
    \begin{aligned}
        \label{eq:SM_Lagrangian6}
        \phi(\boldsymbol{\Pi}) = &\left(\pi_0 + \sum\limits_{k \in \mathcal{K}} \pi_k \right) d 
        - \sum\limits_{i \in \mathcal{I}} \max\limits_{\boldsymbol{x_i} \in \mathcal{X}_i} \Bigg[ 
        \left(\pi_{0} + \sum\limits_{k \in \mathcal{K}} \pi_k - c_i \right) g_{i0} \\
        & + \left(\sum\limits_{k \in \mathcal{K}} \pi_k - q_i^{g,up} \right) r_i^{g,up}  
        - \sum\limits_{k \in \mathcal{K}^{OFF}_i} \pi_k \left(g_{i0} + r_i^{g,up}\right)\Bigg]
    \end{aligned}
\end{equation}

The first term within the square brackets represents the net revenue of generator $i$ from selling energy. Note that, at the optimal solution, energy is priced at $\pi^*_0 + \sum\limits_{k \in \mathcal{K}}\pi^*_k$, which is identical to the price charged to consumers (first term in the right-hand side of \eqref{eq:SM_Lagrangian6}). Moreover, this price is identical to that derived in  \citet{Arroyo_2005} (Table \ref{tab:SM_pricing_system_arroyo}). The second term within the square brackets represents the net revenue from up-spinning reserve, where such a commodity is priced at $\sum\limits_{k \in \mathcal{K}}\pi^*_k$, which is consistent with the security price defined in \citet{Arroyo_2005} (Table \ref{tab:SM_pricing_system_arroyo}). The last term in \eqref{eq:SM_Lagrangian6} is here coined as ``security charge'' and can be viewed as what generator $i$ should pay back for being responsible for system required reserves. Importantly, energy and reserve price incentives are uniform across all generators, whereas the security charge provides specific incentives to generators based on their contribution to the system-wide reserve needs.

Therefore, different from the settlement described in \citet{Arroyo_2005}, the total revenue for each generator consists of three components: the energy revenue (energy price times pre-contingency power output), the up-spinning reserve revenue (up-spinning reserve price times up-spinning reserve), and the security charge (sum of the corresponding post-contingency Lagrange multipliers times the sum of pre-contingency power output and up-spinning reserve). Table \ref{tab:SM_pricing_system} shows the proposed pricing scheme and the corresponding settlement.

\begin{table}[h!]
\centering
\captionsetup{justification=centering, width=\linewidth}
\caption{Proposed pricing system and settlement for problem \eqref{eq:SMobj}--\eqref{eq:SMcon6}}
\label{tab:SM_pricing_system}
\resizebox{0.5\textwidth}{!}{ 
\begin{tabular}{|l|l|}
\hline
\multirow{2}{*}{\textbf{Energy Price}} &
  \multirow{2}{*}{$ p^e = \pi^*_0 + \sum\limits_{k \in \mathcal{K}} \pi^*_k$} \\
                                                    &                                             \\ \hline
\multirow{2}{*}{\textbf{\begin{tabular}[c]{@{}l@{}}Up-Spinning \\ Reserve Price\end{tabular}}} &
  \multirow{2}{*}{$p^{up} = \sum\limits_{k \in \mathcal{K}} \pi^*_k $} \\
                                                    &                                             \\ \hline
\multirow{2}{*}{\textbf{Generation Energy Revenue}} & \multirow{2}{*}{$R^{g,e}_i = p^e g_{i0}$}       \\
                                                    &                                             \\ \hline
\multirow{2}{*}{\textbf{\begin{tabular}[c]{@{}l@{}}Generation Up-Spinning\\ Reserve Revenue\end{tabular}}} &
  \multirow{2}{*}{$R^{g,up}_i = p^{up} r^{g,up}_{i}$} \\
                                                    &                                             \\ \hline
\multirow{2}{*}{\textbf{Generation Security Charge}} &
  \multirow{2}{*}{$C^{g,s}_i = \sum\limits_{k \in \mathcal{K}^{OFF}_i}\pi^*_k\left(g_{i0} + r^{g,up}_{i} \right)$} \\
                                                    &                                             \\ \hline
\multirow{2}{*}{\textbf{Generation Total Revenue}} &
  \multirow{2}{*}{$ R^{g,t}_i = R^{g,e}_i + R^{g,up}_i - C^{g,s}_i$} \\
                                                    &                                             \\ \hline
\multirow{2}{*}{\textbf{Consumer Payment}}          & \multirow{2}{*}{$CP = p^e d $}              \\
                                                    &                                             \\ \hline
\multirow{2}{*}{\textbf{Generation Energy Cost}}    & \multirow{2}{*}{$C^{g,e}_i = c_ig_{i0}$}        \\
                                                    &                                             \\ \hline
\multirow{2}{*}{\textbf{\begin{tabular}[c]{@{}l@{}}Generation Up-Spinning \\ Reserve Cost\end{tabular}}} &
  \multirow{2}{*}{$C^{g,up}_i = q_i^{g,up}r_i^{g,up}$} \\
                                                    &                                             \\ \hline
\multirow{2}{*}{\textbf{Generation Total Cost}}     & \multirow{2}{*}{$C^{g,t}_i = C^{g,e}_i + C^{g,up}_i$}   \\
                                                    &                                             \\ \hline
\multirow{2}{*}{\textbf{Generation Profit}}         & \multirow{2}{*}{$Profit^g_i = R^{g,t}_i - C^{g,t}_i$} \\
                                                    &                                             \\ \hline
\end{tabular}
}
\end{table}

The proposed pricing system has been applied to the illustrative example described in Section \ref{sect: sm_example_arroyo}. Using the results provided in Tables \ref{tab:schedules_1bus} and \ref{tab:lagrange_1bus} in the expressions listed in Table \ref{tab:SM_pricing_system} gives rise to the results reported in Table \ref{tab:revenues_1bus_3Gen}, which lists the breakdown of revenues, costs, charges, and profits for the three generators. Compared to the results reported in Table \ref{tab:revenues_1bus_3Gen_arroyo}, the revenues from energy and reserves and the associated costs are identical. However, as a major salient result, generator 1 is subject to a security charge under the proposed pricing scheme. Note that this generator is the primary driver of the reserve requirement, as indicated by the non-zero Lagrange multiplier in Table \ref{tab:lagrange_1bus}. This charge accounts for the increased system costs associated with the need for up-spinning reserves. Specifically, generator 1's failure requires $65$ MW of up-spinning reserve.  As a result, under the proposed method, the total generation revenue is $\$12,000$, which exactly matches the consumer payment (Table \ref{tab:system_settl_1bus_LR}), thereby overcoming the issue featured by the approach described in \citet{Arroyo_2005}. As can be seen in Table \ref{tab:revenues_1bus_3Gen}, the consideration of the security charge yields a reduced profit for generator 1 compared to that reported in Table \ref{tab:revenues_1bus_3Gen_arroyo}. Note, however, that the new resulting profit is non-negative. 

\begin{table}[h!]
\captionsetup{justification=centering, width=\linewidth}
\caption{Single-bus example -- Proposed generation settlement (\$)}
\label{tab:revenues_1bus_3Gen}
\centering
\resizebox{0.6\textwidth}{!}{
\renewcommand{\arraystretch}{1.2} 
\begin{tabular}{ccccccccc}
\hline
${i}$ &
  ${R^{g,e}_i}$ &
  ${R^{g,up}_i}$ &
  ${C^{g,s}_i}$ &
  ${R^{g,t}_i}$ &
  ${C^{g,e}_i}$ &
  ${C^{g,up}_i}$ &
  ${C^{g,t}_i}$ &
  ${Profit^g_i}$ \\ \hline
\textbf{1} &
  \textcolor{white}{0}6,500 &
  \textcolor{white}{000,}0 &
  5,200 &
  \textcolor{white}{0}1,300 &
  1,300 &
  \textcolor{white}{00}0 &
  1,300 &
  \textcolor{white}{000,}0 \\
\textbf{2} &
  \textcolor{white}{0}3,000 &
  2,400 &
  \textcolor{white}{000,}0 &
  \textcolor{white}{0}5,400 &
  1,500 &
  150 &
  1,650 &
  3,750 \\
\textbf{3} &
  \textcolor{white}{0}2,500 &
  2,800 &
  \textcolor{white}{000,}0 &
  \textcolor{white}{0}5,300 &
  2,500 &
  350 &
  2,850 &
  2,450 \\ \hline
\textbf{Total} &
  \textbf{12,000} &
  \textbf{5,200} &
  \textbf{5,200} &
  \textbf{12,000} &
  \textbf{5,300} &
  \textbf{500} &
  \textbf{5,800} &
  \textbf{6,200} \\ \hline
\end{tabular}
}
\end{table}

\begin{table}[h!]
\captionsetup{justification=centering, width=\linewidth}
\caption{Single-bus example -- Proposed system settlement (\$)}
\label{tab:system_settl_1bus_LR}
\centering
\resizebox{0.3\textwidth}{!}{ 
\begin{tabular}{cc}
\hline
\textbf{Generation Revenue} & 12,000 \\
\textbf{Consumer Payment}   & 12,000 \\ \hline
\textbf{Balance}            & \textcolor{white}{0000,}0 \\ \hline
\end{tabular}
}
\end{table}

\FloatBarrier

The following theorems can be set forth for the proposed pricing framework,  their respective proofs being presented in Appendices B and C.

\begin{theorem}
\label{theorem1}
(Revenue Adequacy): The proposed pricing framework ensures that generators’ profits are non-negative.
\end{theorem}

\begin{theorem}
\label{theorem2}
(Revenue Neutrality): There is no missing money in the market, ensuring the efficient alignment between the consumer payment and the total generation revenue. 
\end{theorem}

\section{Separately Pricing Up and Down Reserves and Transmission Pricing}
\label{sect:up and down reserves}

In this section, the scope is broadened to consider a more realistic setting. To that end, network constraints and demand response are now incorporated into the market-clearing model, whereas both generation and line outages are accounted for. As a consequence, the set of ancillary services is extended to also include downward reserves, which are co-optimized with energy and upward reserves. This model is useful to comprehensively present the salient features of the proposed pricing scheme compared to that in \citet{Arroyo_2005}, which, aside from the concept of security charges already discussed in Section 2, comprise 1) separately pricing upward and downward reserves, and 2) transmission pricing.

\subsection{Optimization Model}
The market-clearing model is formulated as the following linear program:

\allowdisplaybreaks
\vspace{-3.5ex}
\begin{align}
& \min\limits_{\substack{\boldsymbol{x} \geq \boldsymbol{0}, \boldsymbol{y} \geq \boldsymbol{0} \\ \boldsymbol{\theta_0}, \boldsymbol{\theta_k}}} \quad 
\boldsymbol{c^{\top}g_0} + (\boldsymbol{q^{g,up}})^{\top} \boldsymbol{r^{g,up}}  + (\boldsymbol{q^{g,dn}})^{\top} \boldsymbol{r^{g,dn}} - \boldsymbol{w^{\top} d_0} \nonumber \\ 
& \quad \quad \quad + (\boldsymbol{q^{d,up}})^{\top} \boldsymbol{r^{d,up}} + (\boldsymbol{q^{d,dn}})^{\top} \boldsymbol{r^{d,dn}} \label{eq:EMobj} \\   
& \text{subject to:} &  \nonumber \\
& \boldsymbol{M^gg_0 + AH\theta_0} = \boldsymbol{M^dd_0} \quad \quad \   (\boldsymbol{\pi_0}) \label{eq:EMcon2} \\
&\boldsymbol{M^gg_k + A_kH_k\theta_k} = \boldsymbol{M^dd_k} \quad \  
 (\boldsymbol{\pi_k}), \quad \quad \quad  \quad  \quad \forall k \in \mathcal{K} \label{eq:EMcon3} \\
& \boldsymbol{-F} \leq \boldsymbol{H\theta_0} \leq \boldsymbol{F}   \quad  \quad \quad  \quad  \quad \quad   \left(\boldsymbol{\pi_0^{f+}, \pi_0^{f-}}\right) \label{eq:EMcon4}\\
&-\boldsymbol{F} \leq  \boldsymbol{H_k\theta_k} \leq\boldsymbol{F} \quad \quad \quad \quad \quad  \left(\boldsymbol{\pi_k^{f+}, \pi_k^{f-}}\right), \quad \ \forall k \in \mathcal{K} \label{eq:EMcon5} \\
&\boldsymbol{g_k} \leq \mathbb{D}(\boldsymbol{a^g_k})(\boldsymbol{ g_0+r^{g,up}}), \quad \quad \quad \quad \quad \quad \quad \quad \quad \quad  \forall k \in \mathcal{K}  \label{eq:EMcon7} \\
&\boldsymbol{g_k} \geq  \mathbb{D}(\boldsymbol{a^g_k})(\boldsymbol{g_0-r^{g,dn}}), \quad \quad \quad \quad \quad \quad \quad \quad \quad \ \ \ \forall k \in \mathcal{K} \label{eq:EMcon6} \\
&\boldsymbol{g_0 + r^{g,up}} \leq \boldsymbol{\overline{G}} \label{eq:EMcon8}\\
&\boldsymbol{g_0 - r^{g,dn}} \geq \boldsymbol{0} \label{eq:EMcon9}\\
&  \boldsymbol{r^{g,up}}  \leq \boldsymbol{\overline{R}^{g,up}} \label{eq:EMRESDN} \\
&  \boldsymbol{r^{g,dn}}  \leq \boldsymbol{\overline{R}^{g,dn}} \label{eq:EMRESDN} \\
&\boldsymbol{d_k} \geq \boldsymbol{ d_0-r^{d,up} }, \quad \quad \quad \quad \quad \quad \quad \quad \quad \quad \quad \quad  \ \forall k \in \mathcal{K}\ \label{eq:EMcon10a} \\
&\boldsymbol{d_k} \leq \boldsymbol{ d_0+r^{d,dn} }, \quad \quad \quad \quad \quad \quad \quad \quad \quad \quad \quad \quad   \ \forall k \in \mathcal{K}\ \label{eq:EMcon10}\\
&\boldsymbol{d_0 - r^{d,up}} \geq \boldsymbol{0} \label{eq:EMcon11}\\
&\boldsymbol{d_0 + r^{d,dn}} \leq \boldsymbol{\overline{D}}\label{eq:EMcon12}\\
&\boldsymbol{r^{d,up}} \leq \boldsymbol{\overline{R}^{d,up}}  \label{eq:EMcon13.2}\\
&\boldsymbol{r^{d,dn}} \leq \boldsymbol{\overline{R}^{d,dn}}  \label{eq:EMcon13}
\end{align}

\noindent
where $\boldsymbol{x} = \{\boldsymbol{g_0}, \boldsymbol{r^{g,up}}, \boldsymbol{r^{g,dn}}, \{\boldsymbol{g_k}\}_{k\in\mathcal{K}} \}$ and  $\boldsymbol{y} = \{ \boldsymbol{d_0}, \boldsymbol{r^{d,up}}, \boldsymbol{r^{d,dn}}, \{\boldsymbol{d_k}\}_{k\in\mathcal{K}}  \}$. 

As per \eqref{eq:EMobj}, the optimization goal is the minimization of the sum of the costs for generating power and providing up- and down-spinning reserves offered by the generators minus the sum of the bid utility functions for consuming power plus the costs for providing up- and down-spinning reserves by the loads. 

Using a dc load flow model, constraints \eqref{eq:EMcon2} and \eqref{eq:EMcon3} represent the nodal power balance equations under the pre-contingency and contingency states, respectively. Additionally, constraints \eqref{eq:EMcon4} and \eqref{eq:EMcon5} enforce the line flow capacity limits. The corresponding Lagrange multipliers are shown in parentheses.

Constraints \eqref{eq:EMcon7} and \eqref{eq:EMcon6} relate the up- and down-spinning reserve contributions from generators to the production levels under the pre-contingency and contingency states. Furthermore, constraints \eqref{eq:EMcon8}--\eqref{eq:EMRESDN} ensure that such generation and reserve levels remain within their respective operational boun\-daries.

Constraints \eqref{eq:EMcon10a} and \eqref{eq:EMcon10} establish the relationship between the up- and down-spinning reserve contributions from consumers and the power levels consumed under the pre-contingency and contingency states. Finally, demand-related bounds are set in constraints \eqref{eq:EMcon11}--\eqref{eq:EMcon13}.

\subsection{Challenges of the Existing Pricing Framework}
\label{sect: em_example_arroyo}

According to \citet{Arroyo_2005}, nodal prices for energy and security can be defined for problem  \eqref{eq:EMobj}--\eqref{eq:EMcon13}, as summarized in Table \ref{tab:EM_pricing_system_arroyo}. The corresponding generation and consumer settlements are provided in Tables \ref{tab:EM_pricing_system_arroyo2} and \ref{tab:EM_pricing_system_arroyo3}, respectively. As can be seen, for each bus, a single price, namely the nodal security price, is used for all spinning reserves provided by the generators and consumers at that bus.

\begin{table}[h!]
\centering
\captionsetup{justification=centering, width=\linewidth}
\caption{Nodal prices for problem \eqref{eq:EMobj}--\eqref{eq:EMcon13} according to the pricing system of \citet{Arroyo_2005}}
\label{tab:EM_pricing_system_arroyo}
\resizebox{0.4\textwidth}{!}{ 
\begin{tabular}{|l|l|}
\hline
\multirow{2}{*}{\textbf{Nodal Energy Price}}   & \multirow{2}{*}{$p^e_b = \pi^*_{b0} + \sum\limits_{k \in \mathcal{K}}\pi^*_{bk}$} \\
                                               &                                                                               \\ \hline
\multirow{2}{*}{\textbf{Nodal Security Price}} & \multirow{2}{*}{$p_b^{s} = \sum\limits_{k \in \mathcal{K}}\pi^*_{bk}$}          \\
                                               &                                                                               \\ \hline
\end{tabular}
}
\end{table}

\begin{table}[h!]
\centering
\captionsetup{justification=centering, width=\linewidth}
\caption{Generation settlement for problem \eqref{eq:EMobj}--\eqref{eq:EMcon13} according to the pricing system of \citet{Arroyo_2005}}
\label{tab:EM_pricing_system_arroyo2}
\resizebox{0.5\textwidth}{!}{ 
\begin{tabular}{|l|l|}
\hline
\multirow{2}{*}{\textbf{Generation Energy Revenue}} &
  \multirow{2}{*}{$R^{g,e}_i = p^e_{b{(i)}}g_{i0}$} \\
 &
   \\ \hline
\multirow{2}{*}{\textbf{\begin{tabular}[c]{@{}l@{}}Generation Up-Spinning \\ Reserve Revenue\end{tabular}}} &
  \multirow{2}{*}{$R^{g,up}_i = p^{s}_{b(i)}r^{g,up}_{i}$} \\
 &
   \\ \hline
\multirow{2}{*}{\textbf{\begin{tabular}[c]{@{}l@{}}Generation Down-Spinning \\ Reserve Revenue\end{tabular}}} &
  \multirow{2}{*}{$R^{g,dn}_i = p^{s}_{b(i)}r^{g,dn}_i$} \\
 &
   \\ \hline
\multirow{2}{*}{\textbf{Generation Total Revenue}} &
  \multirow{2}{*}{$R^{g,t}_i = R^{g,e}_i + R^{g,up}_i + R^{g,dn}_i$} \\
 &
   \\ \hline
\multirow{2}{*}{\textbf{Generation Energy Cost}} &
  \multirow{2}{*}{$C^{g,e}_i = c_ig_{i0}$} \\
 &
   \\ \hline
\multirow{2}{*}{\textbf{\begin{tabular}[c]{@{}l@{}}Generation Up-Spinning \\ Reserve Cost\end{tabular}}} &
  \multirow{2}{*}{$C^{g,up}_i = q_i^{g,up}r_i^{g,up}$} \\
 &
   \\ \hline
\multirow{2}{*}{\textbf{\begin{tabular}[c]{@{}l@{}}Generation Down-Spinning\\ Reserve Cost\end{tabular}}} &
  \multirow{2}{*}{$C^{g,dn}_i = q_i^{g,dn}r_i^{g,dn}$} \\
 &
   \\ \hline
\multirow{2}{*}{\textbf{Generation Total Cost}} &
  \multirow{2}{*}{$C^{g,t}_i = C^{g,e}_i + C^{g,up}_i + C^{g,dn}_i $} \\
 &
   \\ \hline
\multirow{2}{*}{\textbf{Generation Profit}} &
  \multirow{2}{*}{$Profit^g_i = R^{g,t}_i - C^{g,t}_i$} \\
 &
   \\ \hline
\end{tabular}
}
\end{table}

\begin{table}[h!]
\centering
\captionsetup{justification=centering, width=\linewidth}
\caption{Consumer settlement for problem \eqref{eq:EMobj}--\eqref{eq:EMcon13} according to the pricing system of \citet{Arroyo_2005}}
\label{tab:EM_pricing_system_arroyo3}
\resizebox{0.5\textwidth}{!}{ 
\begin{tabular}{|l|l|}
\hline
\multirow{2}{*}{\textbf{Consumer Payment for Energy}} & \multirow{2}{*}{$CP_j^e = p^e_{b(j)}d_{j0}$}               \\
                                                      &                                                            \\ \hline
\multirow{2}{*}{\textbf{\begin{tabular}[c]{@{}l@{}}Consumer Up-Spinning \\ Reserve Revenue\end{tabular}}}  & \multirow{2}{*}{$R^{d,up}_j = p^{s}_{b(j)}r^{d,up}_j$} \\
                                                      &                                                            \\ \hline
\multirow{2}{*}{\textbf{\begin{tabular}[c]{@{}l@{}}Consumer Down-Spinning\\ Reserve Revenue\end{tabular}}} & \multirow{2}{*}{$R^{d,dn}_j = p^{s}_{b(j)}r^{d,dn}_j$} \\
                                                      &                                                            \\ \hline
\multirow{2}{*}{\textbf{Consumer Utility}}            & \multirow{2}{*}{$U^d_j = w_jd_{j0}$}                       \\
                                                      &                                                            \\ \hline
\multirow{2}{*}{\textbf{\begin{tabular}[c]{@{}l@{}}Consumer Up-Spinning \\ Reserve Cost\end{tabular}}}     & \multirow{2}{*}{$C^{d,up}_j = q_j^{d,up}r_j^{d,up}$}   \\
                                                      &                                                            \\ \hline
\multirow{2}{*}{\textbf{\begin{tabular}[c]{@{}l@{}}Consumer Down-Spinning\\ Reserve Cost\end{tabular}}}    & \multirow{2}{*}{$C^{d,dn}_j = q_j^{d,dn}r_j^{d,dn}$}   \\
                                                      &                                                            \\ \hline
\multirow{2}{*}{\textbf{Consumer Total  Cost}}        & \multirow{2}{*}{$C^{d,t}_j = C^{d,up}_j+C^{d,dn}_j$}     \\
                                                      &                                                            \\ \hline
\multirow{2}{*}{\textbf{Consumer Payment}}            & \multirow{2}{*}{$CP_j = CP_j^e - R^{d,up}_j - R^{d,dn}_j$} \\
                                                      &                                                            \\ \hline
\multirow{2}{*}{\textbf{Consumer Profit}}             & \multirow{2}{*}{$Profit^d_j = U^d_j - C^{d,t}_j - CP_j $}            \\
                                                      &                                                            \\ \hline
\end{tabular}
}
\end{table}

\begin{figure}[!ht]
\centering
\includegraphics[scale = .45]{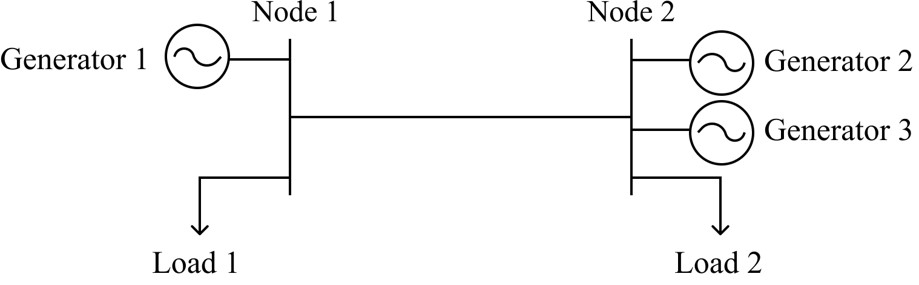}
\caption{{Two-bus example -- One-line diagram.}}
\label{fig:2bus}
\end{figure}

For illustration purposes, the pricing system presented in Table \ref{tab:EM_pricing_system_arroyo} is applied to the two-bus, one-line system depicted in Figure \ref{fig:2bus}, where the generation fleet is based on that considered in Section \ref{sect: sm_example_arroyo}. The line reactance is $1$ p.u. on a base of $100$ MVA and $138$ kV, whereas the line capacity is $70$ MVA. In Table \ref{tab:gen_data_2bus}, the generation data presented in Table \ref{tab:gen_data_1bus} are extended with the information for down-spinning reserve offers, which is the same as for up-spinning reserves. Load data are provided in Table  \ref{tab:load_data_2bus}. The analysis considers four credible contingencies: the failure of each generator and the outage of the line.

Problem \eqref{eq:EMobj}--\eqref{eq:EMcon13} has been solved to optimality using the simplex algorithm of CPLEX. Table \ref{tab:schedules_gen_2bus} displays the optimal results, which feature a total value of the objective function equal to $-\$15,475$. In other words, the optimal social welfare amounts to $\$15,475$. It should be noted that the optimal pre-contingency system demand amounts to 120 MW, which is identical to the value of the inelastic demand in the single-bus example. Consequently, the optimal generation schedule is similar to that attained for the single-bus example, the main difference being the $5$ MW of down-spinning reserve scheduled to generator 2. It is also worth highlighting that load 1 is scheduled to provide 
upward spinning reserve, which is deployed in response to contingencies involving the loss of generator 1 or line 1--2.

\begin{table}[h!]
\caption{Two-bus example -- Generation data}
\label{tab:gen_data_2bus}
\centering
\resizebox{0.5\textwidth}{!}{
\renewcommand{\arraystretch}{1.2} 
\begin{tabular}{ccccccc}
\hline
\textbf{$i$} &
  \multicolumn{1}{c}{\begin{tabular}[c]{@{}c@{}}$\overline{G}_i$\\ (MW)\end{tabular}} &
  \multicolumn{1}{c}{\begin{tabular}[c]{@{}c@{}}$\overline{R}^{g,up}_i$\\ (MW)\end{tabular}} &
  \multicolumn{1}{c}{\begin{tabular}[c]{@{}c@{}}$\overline{R}^{g,dn}_i$\\ (MW)\end{tabular}} &
  \multicolumn{1}{c}{\begin{tabular}[c]{@{}c@{}}$c_i$\\ (\$/MWh)\end{tabular}} &
  \multicolumn{1}{c}{\begin{tabular}[c]{@{}c@{}}$q_i^{g,up}$\\ (\$/MW)\end{tabular}} &
  \multicolumn{1}{c}{\begin{tabular}[c]{@{}c@{}}$q_i^{g,dn}$\\ (\$/MW)\end{tabular}} \\ \hline
\textbf{1} & 100 & 50 & 50 & \textcolor{white}{0}20  & \textcolor{white}{0}2  & \textcolor{white}{0}2 \\
\textbf{2} & \textcolor{white}{0}60  & 30 & 30 & \textcolor{white}{0}50  & \textcolor{white}{0}5  & \textcolor{white}{0}5 \\
\textbf{3} & \textcolor{white}{0}70  & 35 & 35 & 100 & 10 & 10 \\ \hline
\end{tabular}
}
\end{table}

\begin{table}[h!]
\caption{Two-bus example -- Load data}
\label{tab:load_data_2bus}
\centering
\resizebox{0.5\textwidth}{!}{
\renewcommand{\arraystretch}{1.2} 
\begin{tabular}{ccccccc}
\hline
\textbf{$j$} &
  \begin{tabular}[c]{@{}c@{}}$\overline{D}_j$\\ (MW)\end{tabular} &
  \begin{tabular}[c]{@{}c@{}}$\overline{R}_j^{d,up}$\\ (MW)\end{tabular} &
  \begin{tabular}[c]{@{}c@{}}$\overline{R}_j^{d,dn}$\\ (MW)\end{tabular} &
  \begin{tabular}[c]{@{}c@{}}$w_j$\\ (\$/MWh)\end{tabular} &
  \begin{tabular}[c]{@{}c@{}}$q^{d,up}_j$\\ (\$/MW)\end{tabular} &
  \begin{tabular}[c]{@{}c@{}}$q^{d,dn}_j$\\ (\$/MW)\end{tabular} \\ \hline
\textbf{1} &
  90 &
  10 &
  10 &
  200 &
  150 &
  300 \\
\textbf{2} &
  40 &
  10 &
  10 &
  150 &
  100 &
  250 \\ \hline
\end{tabular}
}
\end{table}

\begin{table}[h!]
\caption{Two-bus example -- Optimal results (MW)}
\label{tab:schedules_gen_2bus}
\centering
\resizebox{0.8\textwidth}{!}{
\renewcommand{\arraystretch}{1.2} 
\begin{tabular}{cccccccc}
\hline
 &
   &
   &
   &
  \multicolumn{4}{c}{\textbf{Generation under Contingency}} \\ \cline{5-8} 
\textbf{Generator} &
  \textbf{\begin{tabular}[c]{@{}c@{}}Pre-Contingency\\ Generation\end{tabular}} &
  \textbf{\begin{tabular}[c]{@{}c@{}}Up-Spinning \\ Reserve\end{tabular}} &
  \textbf{\begin{tabular}[c]{@{}c@{}}Down-Spinning \\ Reserve\end{tabular}} &
  \textbf{\begin{tabular}[c]{@{}c@{}}Outage of\\ Generator 1\end{tabular}} &
  \textbf{\begin{tabular}[c]{@{}c@{}}Outage of\\ Generator 2\end{tabular}} &
  \textbf{\begin{tabular}[c]{@{}c@{}}Outage of\\ Generator 3\end{tabular}} &
  \textbf{\begin{tabular}[c]{@{}c@{}}Outage of\\ Line 1--2\end{tabular}} \\ \hline
\textbf{1} &
  \textcolor{white}{0}75 &
  \textcolor{white}{0}0 &
  0 &
  \textcolor{white}{00}0 &
  \textcolor{white}{0}75 &
  \textcolor{white}{0}75 &
  \textcolor{white}{0}75 \\
\textbf{2} &
  \textcolor{white}{0}30 &
  30 &
  5 &
  \textcolor{white}{0}60 &
  \textcolor{white}{00}0 &
  \textcolor{white}{0}45 &
  \textcolor{white}{0}25 \\
\textbf{3} &
  \textcolor{white}{0}15 &
  35 &
  0 &
  \textcolor{white}{0}50 &
  \textcolor{white}{0}45 &
  \textcolor{white}{00}0 &
  \textcolor{white}{0}15 \\
\textbf{Total} &
  \textbf{120} &
  \textbf{65} &
  \textbf{5} &
  \textbf{110} &
  \textbf{120} &
  \textbf{120} &
  \textbf{115} \\ \hline
 &
   &
   &
   &
  \multicolumn{4}{c}{\textbf{Demand under Contingency}} \\ \cline{5-8} 
\textbf{Load} &
  \textbf{\begin{tabular}[c]{@{}c@{}}Pre-Contingency\\ Demand\end{tabular}} &
  \textbf{\begin{tabular}[c]{@{}c@{}}Up-Spinning \\ Reserve\end{tabular}} &
  \textbf{\begin{tabular}[c]{@{}c@{}}Down-Spinning \\ Reserve\end{tabular}} &
  \textbf{\begin{tabular}[c]{@{}c@{}}Outage of\\ Generator 1\end{tabular}} &
  \textbf{\begin{tabular}[c]{@{}c@{}}Outage of\\ Generator 2\end{tabular}} &
  \textbf{\begin{tabular}[c]{@{}c@{}}Outage of\\ Generator 3\end{tabular}} &
  \textbf{\begin{tabular}[c]{@{}c@{}}Outage of\\ Line 1--2\end{tabular}} \\ \hline
\textbf{1} &
  \textcolor{white}{0}80 &
  10 &
  0 &
  \textcolor{white}{0}70 &
  \textcolor{white}{0}80 &
  \textcolor{white}{0}80 &
  \textcolor{white}{0}75 \\
\textbf{2} &
  \textcolor{white}{0}40 &
  \textcolor{white}{0}0 &
  0 &
  \textcolor{white}{0}40 &
  \textcolor{white}{0}40 &
  \textcolor{white}{0}40 &
  \textcolor{white}{0}40 \\
\textbf{Total} &
  \textbf{120} &
  \textbf{10} &
  \textbf{0} &
  \textbf{110} &
  \textbf{120} &
  \textbf{120} &
  \textbf{115} \\ \hline
 &
   &
   &
   &
  \multicolumn{4}{c}{\textbf{Power Flow 1--2 under Contingency}} \\ \cline{5-8} 
\textbf{Line} &
  \textbf{\begin{tabular}[c]{@{}c@{}}Pre-Contingency\\ Power Flow 1--2\end{tabular}} &
  \textbf{} &
  \textbf{} &
  \textbf{\begin{tabular}[c]{@{}c@{}}Outage of\\ Generator 1\end{tabular}} &
  \textbf{\begin{tabular}[c]{@{}c@{}}Outage of\\ Generator 2\end{tabular}} &
  \textbf{\begin{tabular}[c]{@{}c@{}}Outage of\\ Generator 3\end{tabular}} &
  \textbf{\begin{tabular}[c]{@{}c@{}}Outage of\\ Line 1--2\end{tabular}} \\ \hline
\textbf{1--2} &
  -5 &
   &
   &
  -70 &
  -5 &
  -5 &
  0 \\ \hline
\end{tabular}
}
\end{table}

The outage of generator 1 constitutes a critical contingency for the system, requiring the full deployment of the up-spinning reserves scheduled to generators 2 and 3 and load 1. Analogously, the line outage is another critical contingency that compels generator 2 to reduce its output by fully utilizing its down-spinning reserve. Note, however, that under this contingency state a marginal increase in load 2 would actually reduce the need for down-spinning reserves at this bus. This explains why the corresponding Lagrange multiplier is negative (Table \ref{tab:lagrange_bus_2bus}). Finally, it is worth pointing out that line congestion is solely experienced during the outage of generator 1. In fact, the pre-contingency power flow is well below the line capacity to allow generators 2 and 3 to ramp up during this critical outage. 

\begin{table}[h!]
\caption{Two-bus example -- Lagrange multipliers for the nodal power balance equations (\$/MWh)}
\label{tab:lagrange_bus_2bus}
\centering
\resizebox{0.6\textwidth}{!}{ 
\begin{tabular}{cccccc}
\hline
\textbf{Bus} &
  \textbf{\begin{tabular}[c]{@{}c@{}}Pre-Contingency\\ State\end{tabular}} &
  \textbf{\begin{tabular}[c]{@{}c@{}}Outage of \\ Generator 1\end{tabular}} &
  \textbf{\begin{tabular}[c]{@{}c@{}}Outage of\\ Generator 2\end{tabular}} &
  \textbf{\begin{tabular}[c]{@{}c@{}}Outage of\\ Generator 3\end{tabular}} &
  \textbf{\begin{tabular}[c]{@{}c@{}}Outage of\\ Line 1--2\end{tabular}} \\ \hline
\textbf{1} &
  20 &
  180 &
  0 &
  0 &
  \textcolor{white}{-}0 \\
\textbf{2} &
  20 &
  \textcolor{white}{0}85 &
  0 &
  0 &
  -5 \\ \hline
\end{tabular}
}
\end{table}

The nodal prices and the settlement for generators and consumers calculated according to Tables \ref{tab:EM_pricing_system_arroyo}, \ref{tab:EM_pricing_system_arroyo2}, and \ref{tab:EM_pricing_system_arroyo3} are presented in Tables \ref{tab:prices_2bus_Arroyo},  \ref{tab:revenues_gen_2bus_Arroyo}, and \ref{tab:revenues_consumers_2bus_Arroyo}, respectively. As can be seen, the profit distribution among market agents is economically inconsistent as the total profit of generators and consumers is $44.6\%$ greater than the optimal social welfare. Furthermore, similar to the previous example, the pricing scheme proposed in \citet{Arroyo_2005} results in total generation revenue exceeding total consumer payment, as summarized in Table \ref{tab:system_settl_2bus}. This issue becomes more important in a network-constrained setting because the absence of a transmission surplus to compensate for lines creates an additional financial burden. For this particular example, the consumer payment is insufficient to cover the generation revenue and the transmission revenue associated with the line congestion under the outage of generator 1. The resulting settlement imbalance may be addressed by out-of-market compensation that may give rise to discrimination among market participants and inadequate economic signals, among other issues. Alternatively, in Section \ref{sect:pricing_system_EM} we propose a new pricing scheme that does not feature settlement imbalance while relying on a sound mathematical framework.

\begin{table}[h!]
\caption{Two-bus example -- Nodal prices according to \citet{Arroyo_2005} (\$/MWh)}
\label{tab:prices_2bus_Arroyo}
\centering
\resizebox{0.4\textwidth}{!}{
\begin{tabular}{ccc}
\hline
\textbf{$b$} &
  \multicolumn{1}{c}{\textbf{\begin{tabular}[c]{@{}c@{}}Nodal Energy Price, \\ $p^e_b$\end{tabular}}} &
  \multicolumn{1}{c}{\textbf{\begin{tabular}[c]{@{}c@{}}Nodal Security Price,\\ $p^s_b$\end{tabular}}} \\ \hline
\textbf{1} &
  200 &
  180 \\
\textbf{2} &
  100 &
  \textcolor{white}{0}80 \\ \hline
\end{tabular}
}
\end{table}

\begin{table}[h!]
\captionsetup{justification=centering, width=\linewidth}
\caption{Two-bus example -- Generation settlement according to \citet{Arroyo_2005} (\$)}
\label{tab:revenues_gen_2bus_Arroyo}
\centering
\resizebox{0.6\textwidth}{!}{ 
\renewcommand{\arraystretch}{1.2} 
\begin{tabular}{cccccccccc}
\hline
$i$ &
  $R^{g,e}_i$ &
  $R_i^{g,up}$ &
  $R_i^{g,dn}$ &
  $R^{g,t}_i$ &
  $C^{g,e}_i$ &
  $C^{g,up}_i$ &
  $C^{g,dn}_i$ &
  $C^{g,t}_i$ &
  $Profit^g_i$ \\ \hline
\textbf{1} &
  15,000 &
  \textcolor{white}{000,}0 &
  \textcolor{white}{00}0 &
  15,000 &
  1,500 &
  \textcolor{white}{00}0 &
  \textcolor{white}{0}0 &
  1,500 &
  13,500 \\
\textbf{2} &
  \textcolor{white}{0}3,000 &
  2,400 &
  400 &
  \textcolor{white}{0}5,800 &
  1,500 &
  150 &
  25 &
  1,675 &
  \textcolor{white}{0}4,125 \\
\textbf{3} &
  \textcolor{white}{0}1,500 &
  2,800 &
  \textcolor{white}{00}0 &
  \textcolor{white}{0}4,300 &
  1,500 &
  350 &
  \textcolor{white}{0}0 &
  1,850 &
  \textcolor{white}{0}2,450 \\ \hline
\textbf{Total} &
  \textbf{19,500} &
  \textbf{5,200} &
  \textbf{400} &
  \textbf{25,100} &
  \textbf{4,500} &
  \textbf{500} &
  \textbf{25} &
  \textbf{5,025} &
  \textbf{20,075} \\ \hline
\end{tabular}
}
\end{table}

\begin{table}[h!]
\captionsetup{justification=centering, width=\linewidth}
\caption{Two-bus example -- Consumer settlement according to \citet{Arroyo_2005} (\$)}
\label{tab:revenues_consumers_2bus_Arroyo}
\centering
\resizebox{0.6\textwidth}{!}{ 
\renewcommand{\arraystretch}{1.2} 
\begin{tabular}{cccccccccc}
\hline
$ j$       & $CP^e_j$ & $R^{d,up}_j$ & $R^{d,dn}_j$ & $CP_j$ & $U^d_{j}$ & $C^{d,up}_j$ & $C^{d,dn}_j$ & $C^{d,t}_j$ & $Profit^d_j$             \\ \hline
\textbf{1} & 16,000   & 1,800        & 0            & 14,200 & 16,000    & 1,500        & 0            & 1,500       & \textcolor{white}{0,}300 \\
\textbf{2} &
  \textcolor{white}{0}4,000 &
  \textcolor{white}{000}0 &
  0 &
  \textcolor{white}{0}4,000 &
  \textcolor{white}{0}6,000 &
  \textcolor{white}{000,}0 &
  0 &
  \textcolor{white}{000,}0 &
  2,000 \\ \hline
\textbf{Total} &
  \textbf{20,000} &
  \textbf{1,800} &
  \textbf{0} &
  \textbf{18,200} &
  \textbf{22,000} &
  \textbf{1,500} &
  \textbf{0} &
  \textbf{1,500} &
  \textbf{2,300} \\ \hline
\end{tabular}
}
\end{table}

\begin{table}[h!]
\captionsetup{justification=centering, width=\linewidth}
\caption{Two-bus example -- System settlement according to \citet{Arroyo_2005} (\$)}
\label{tab:system_settl_2bus}
\centering
\resizebox{0.3\textwidth}{!}{ 
\begin{tabular}{cc}
\hline
\textbf{Generation Revenue} & \textcolor{white}{-}25,100 \\
\textbf{Consumer Payment}   & \textcolor{white}{-}18,200 \\ \hline
\textbf{Balance}            & \textcolor{white}{0}-6,900 \\ \hline
\end{tabular}
}
\end{table}

\subsection{Proposed Pricing System}
\label{sect:pricing_system_EM}
Following the procedure described in Section \ref{sect: pricing_SM}, we start by building the LD function through the dualization of constraints \eqref{eq:EMcon2}--\eqref{eq:EMcon5}, giving rise to Equation \eqref{eq:EM_Lagrangian1}:

\allowdisplaybreaks
\begin{equation}
\begin{split}
    \label{eq:EM_Lagrangian1}
    \phi\left(\boldsymbol{\Pi}\right) = 
    \min\limits_{\substack{\boldsymbol{x}\in \mathcal{X} \\ \boldsymbol{y}\in \mathcal{Y} \\ \boldsymbol{\theta_0}, \boldsymbol{\theta_k} }}\Bigg\{ 
    & \boldsymbol{c}^{\top} \boldsymbol{g_0} + (\boldsymbol{q^{g,up}})^{\top} \boldsymbol{r^{g,up}} + (\boldsymbol{q^{g,dn}})^{\top} \boldsymbol{r^{g,dn}} \\
    & - \boldsymbol{w}^{\top} \boldsymbol{d_0} + (\boldsymbol{q^{d,up}})^{\top} \boldsymbol{r^{d,up}} + (\boldsymbol{q^{d,dn}})^{\top} \boldsymbol{r^{d,dn}} \\
    & + \boldsymbol{\pi_0}^{\top} \left(\boldsymbol{M^dd_0 - M^g g_0 - AH \theta_0}\right) \\
    & + \sum\limits_{k \in \mathcal{K}} \left[ \boldsymbol{\pi_k}^{\top} \left(\boldsymbol{M^dd_k - M^g g_k - A_kH_k \theta_k}\right) \right] \\
    & + \left(\boldsymbol{\pi_0^{f+}}\right)^{\top} \left(\boldsymbol{-F - H \theta_0}\right)
    +\left(\boldsymbol{\pi_0^{f-}}\right)^{\top} \left(\boldsymbol{F - H \theta_0}\right) \\
    & + \sum\limits_{k \in \mathcal{K}} \Bigg[ \left(\boldsymbol{\pi_k^{f+}}\right)^{\top} \left(\boldsymbol{-F - H_k \theta_k}\right) \\
    & + \left(\boldsymbol{\pi_k^{f-}}\right)^{\top} \left(\boldsymbol{F - H_k \theta_k}\right) \Bigg]
    \Bigg\}
\end{split}
\end{equation}

\noindent
where $\boldsymbol{\Pi} = \left \{ \boldsymbol{\pi_0, \pi_0^{f+}, \pi_0^{f-}}, \left \{ \boldsymbol{\pi_k}, \boldsymbol{\pi_k^{f+}},\boldsymbol{\pi_k^{f-}}\right \}_{k \in \mathcal{K}} \right \}$,  $\mathcal{X}$ represents the feasibility space associated with generation-related constraints $\boldsymbol{x} \geq \boldsymbol{0}$ and  \eqref{eq:EMcon7}--\eqref{eq:EMRESDN}, whereas $\mathcal{Y}$ is the feasibility set corresponding to consumer-related constraints $\boldsymbol{y} \geq \boldsymbol{0}$ and \eqref{eq:EMcon10a}--\eqref{eq:EMcon13}. Rearranging \eqref{eq:EM_Lagrangian1}, we have: 

\vspace{-3ex}
\begin{equation}
\begin{split}
    \label{eq:EM_Lagrangian2}
     \phi\left(\boldsymbol{\Pi}\right) = 
    \min\limits_{\substack{\boldsymbol{x}\in \mathcal{X} \\ \boldsymbol{y}\in \mathcal{Y} \\ \boldsymbol{\theta_0}, \boldsymbol{\theta_k} }}\Bigg\{ 
        & \left(\boldsymbol{c}^{\top} - \boldsymbol{\pi_0}^{\top} \boldsymbol{M^g}\right) \boldsymbol{g_0}
        - \sum\limits_{k \in \mathcal{K}} \boldsymbol{\pi_k}^{\top} \boldsymbol{M^g} \boldsymbol{g_k} 
        + (\boldsymbol{q^{g,up}})^{\top} \boldsymbol{r^{g,up}} \\ &+(\boldsymbol{q^{g,dn}})^{\top} \boldsymbol{r^{g,dn}} 
        + \left(\boldsymbol{\pi_0}^{\top}\boldsymbol{M^d} - \boldsymbol{w}^{\top} \right) \boldsymbol{d_0} \\
        &+ \sum\limits_{k \in \mathcal{K}} \boldsymbol{\pi_k}^{\top}\boldsymbol{M^d} \boldsymbol{d_k} 
        + (\boldsymbol{q^{d,up}})^{\top} \boldsymbol{r^{d,up}} + (\boldsymbol{q^{d,dn}})^{\top} \boldsymbol{r^{d,dn}} \\
        & - \Bigg[\boldsymbol{\pi_0}^{\top} \boldsymbol{A} + \left(\boldsymbol{\pi_0^{f+} + \pi_0^{f-}}\right)^{\top}\Bigg] \boldsymbol{H \theta_0} \\
        & - \sum\limits_{k \in \mathcal{K}} \Bigg[\boldsymbol{\pi_k}^{\top} \boldsymbol{A_k} + \left(\boldsymbol{\pi_k^{f+}}
        + \boldsymbol{\pi_k^{f-}}\right)^{\top} \Bigg] \boldsymbol{H_k \theta_k} \\
        &- \Bigg[\left(\boldsymbol{\pi_0^{f+} - \pi_0^{f-}}\right)^{\top}  
        + \sum\limits_{k \in \mathcal{K}} \left(\boldsymbol{\pi_k^{f+} - \pi_k^{f-}}\right)^{\top}\Bigg] \boldsymbol{F}
    \Bigg\}
\end{split}
\end{equation}

The last term in Equation \eqref{eq:EM_Lagrangian2} is independent of $\boldsymbol{x}$ and $\boldsymbol{y}$, allowing it to be factored out of the minimization problem. Moreover, leveraging the separability of the resulting minimization problem, Equation \eqref{eq:EM_Lagrangian2} can be equivalently rewritten as Equation \eqref{eq:EM_Lagrangian2_separable}:

\begin{equation}
\begin{split}
    \label{eq:EM_Lagrangian2_separable}
    \phi\left(\boldsymbol{\Pi}\right) = & - \Big[\left(\boldsymbol{\pi_0^{f+}} - \boldsymbol{\pi_0^{f-}}\right)^{\top} 
    + \sum\limits_{k \in \mathcal{K}} \left(\boldsymbol{\pi_k^{f+}} - \boldsymbol{\pi_k^{f-}}\right)^{\top} \Big] \boldsymbol{F} \\
    & + \min\limits_{\boldsymbol{x}\in \mathcal{X}}\Big[
    \left(\boldsymbol{c^{\top} - \pi_0^{\top} M^g}\right) \boldsymbol{g_0} 
    - \sum\limits_{k \in \mathcal{K}} \boldsymbol{\pi_k^{\top} M^g g_k} 
    + (\boldsymbol{q^{g,up}})^{\top} \boldsymbol{r^{g,up}} \\
    &+ (\boldsymbol{q^{g,dn}})^{\top} \boldsymbol{r^{g,dn}} \Big] 
    + \min\limits_{\boldsymbol{y}\in \mathcal{Y}}\Big[ 
    \left(\boldsymbol{\pi_0}^{\top}\boldsymbol{M^d} - \boldsymbol{w}^{\top} \right) \boldsymbol{d_0} 
    + \sum\limits_{k \in \mathcal{K}} \boldsymbol{\pi_k}^{\top}\boldsymbol{M^d} \boldsymbol{d_k} \\
    &+ (\boldsymbol{q^{d,up}})^{\top} \boldsymbol{r^{d,up}} 
    + (\boldsymbol{q^{d,dn}})^{\top} \boldsymbol{r^{d,dn}} \Big] \\
    & - \min\limits_{\boldsymbol{\theta_0}, \boldsymbol{\theta_k}}\Bigg\{
    \Big[\boldsymbol{\pi_0}^{\top} \boldsymbol{A} + \left(\boldsymbol{\pi_0^{f+}} + \boldsymbol{\pi_0^{f-}}\right)^{\top} \Big] \boldsymbol{H \theta_0} \\
    & \quad + \sum\limits_{k \in \mathcal{K}} \Big[\boldsymbol{\pi_k^{\top} A_k} 
    + \left(\boldsymbol{\pi_k^{f+}} + \boldsymbol{\pi_k^{f-}}\right)^{\top} \Big] \boldsymbol{H_k \theta_k} 
    \Bigg\}
\end{split}
\end{equation}

For quick reference, the last minimization term in \eqref{eq:EM_Lagrangian2_separable}, related to phase angles, is expressed in a compact way as $\phi_{\theta}\left(\boldsymbol{\Pi}\right)$. In addition, using a component-wise formulation for the other terms making up $\phi\left(\boldsymbol{\Pi}\right)$, Equation \eqref{eq:EM_Lagrangian2_separable} is equivalently cast as:

\begin{equation}
\begin{aligned}
    \label{eq:EM_Lagrangian4_JM}
    \phi\left(\boldsymbol{\Pi}\right) = &- \sum\limits_{l \in \mathcal{L}} \Bigg[\pi_{l0}^{f+} - \pi_{l0}^{f-}  + \sum\limits_{k \in \mathcal{K}} \left(\pi_{lk}^{f+} - \pi_{lk}^{f-} \right) \Bigg] F_l \\
    &- \sum\limits_{i \in \mathcal{I}} \max\limits_{\boldsymbol{x_i} \in \mathcal{X}_i} \Bigg[ \left(\pi_{b(i)0} - c_i\right) g_{i0}
    - q_i^{g,up} r^{g,up}_i - q^{g,dn}_i r^{g,dn}_i \\
    &+ \sum\limits_{k \in \mathcal{K}}\psi^g_{ik}(\pi_{b(i)k}, \boldsymbol{x_i}) \Bigg]
    - \sum\limits_{j \in \mathcal{J}} \max\limits_{\boldsymbol{y_j} \in \mathcal{Y}_j} \Bigg[ \left(w_j - \pi_{b(j)0} \right) d_{j0} \\
    &- q^{d,up}_j r^{d,up}_j - q^{d,dn}_j r^{d,dn}_j - \sum\limits_{k \in \mathcal{K}}\psi^d_{jk}(\pi_{b(j)k}, \boldsymbol{y_j}) \Bigg] - \phi_{\theta}(\boldsymbol{\Pi}),
\end{aligned}
\end{equation}

\noindent
where $\boldsymbol{x_i} = \{g_{i0}, r^{g,up}_i, r^{g,dn}_i \}$ and $\mathcal{X}_i$ is the subset of constraints related to $\boldsymbol{x_i}$, i.e., constraints \eqref{eq:EMcon8}--\eqref{eq:EMRESDN} and $\boldsymbol{x_i} \geq \boldsymbol{0}$. Analogously, $\boldsymbol{y_j} = \{d_{j0}, r^{d,up}_j, r^{d,dn}_j \}$ and $\mathcal{Y}_j$ represents the subset of constraints related to $\boldsymbol{y_j}$, i.e., constraints \eqref{eq:EMcon11}--\eqref{eq:EMcon13} and $\boldsymbol{y_j} \geq \boldsymbol{0}$.
Additionally, $\psi^g_{ik}(\pi_{b(i)k}, \boldsymbol{x_i})$ and $\psi^d_{jk}(\pi_{b(j)k}, \boldsymbol{y_j})$ respectively represent the revenue fraction earned by generator $i$ and the payment fraction of consumer $j$ under contingency $k$ due to their best response against post-contingency-state price $\pi_{bk}$.

$\psi^g_{ik}(\pi_{b(i)k}, \boldsymbol{x_i})$ depends on the generator availability status $a^g_{ik}$ and on the pre-contingency dispatch and reserve schedule $\boldsymbol{x_i}$:

\vspace{-3.5ex}
\begin{align}
& \psi^g_{ik}(\pi_{b(i)k}, \boldsymbol{x}_i) = \max_{g_{ik} \geq 0} \ \pi_{b(i)k} g_{ik} \label{eq:ojbect_function_gen_rev} \\   
& \text{subject to:} &  \nonumber \\
& g_{ik} \leq a^g_{ik} \Bigl( g_{i0} + r_i^{g,up} \Bigl) \label{eq:upper1} \\ 
& g_{ik} \geq a^g_{ik} \Bigl( g_{i0} - r_i^{g,dn} \Bigl)  \label{eq:lower1}  
\end{align}

Analogously, $\psi^d_{jk}(\pi_{b(j)k}, \boldsymbol{y_j})$ depends on the pre-contingency dispatch and reserve schedule $\boldsymbol{y_j}$: 

\vspace{-3.5ex}
\begin{align}
&\psi^d_{jk}(\pi_{b(j)k}, \boldsymbol{y_j}) = \min_{d_{jk} \geq 0} \ \pi_{b(j)k}d_{jk} \label{eq:objective_function_cons_rev} \\ \   
& \text{subject to:} &  \nonumber \\
&d_{jk} \geq d_{j0} - r_j^{d,up}  \label{eq:upper2} \\
&d_{jk} \leq d_{j0} + r_j^{d,dn} \label{eq:lower2}
\end{align}

The optimal solutions to problems \eqref{eq:ojbect_function_gen_rev}--\eqref{eq:lower1} and \eqref{eq:objective_function_cons_rev}--\eqref{eq:lower2} give rise to three possible outcomes for $\psi^g_{ik}(\pi_{b(i)k}, \boldsymbol{x_i})$ and $\psi^d_{jk}(\pi_{b(j)k}, \boldsymbol{y_j})$.

\begin{enumerate}
    \item If $\pi_{bk}$ is positive, $g_{ik}^*$ and $d_{jk}^*$ will be equal to the upper and lower bounds respectively set in \eqref{eq:upper1} and \eqref{eq:upper2}, i.e., all generators and consumers at bus $b$ will be willing to deploy as much up-spinning reserve as possible under contingency $k$. As a result:
    \begin{itemize}
        \item $g^*_{ik} = a^{g}_{ik}\Bigl(g_{i0}+r_i^{g,up} \Bigl)$ and $\psi^g_{ik}(\pi_{b(i)k}, \boldsymbol{x_i}) =  \pi_{b(i)k}a^g_{ik}\Bigl(g_{i0}+r_i^{g,up}\Bigl)$
        \item $d^*_{jk} = d_{j0} - r_j^{d,up}$ and $\psi^d_{jk}(\pi_{b(j)k}, \boldsymbol{y_j}) = \pi_{b(j)k}\Bigl(d_{j0} - r_j^{d,up}\Bigl)$
    \end{itemize}

    \item If $\pi_{bk}$ is negative,  $g_{ik}^*$ and $d_{jk}^*$ will be equal to the lower and upper bounds respectively set in \eqref{eq:lower1} and \eqref{eq:lower2}, i.e., all generators and consumers at bus $b$ will be willing to deploy as much down-spinning reserve as possible under contingency $k$. As a result:

    \begin{itemize}
        \item $g^*_{ik} = a^g_{ik}\Bigl(g_{i0} - r_i^{g,dn} \Bigl)$ and $\psi^g_{ik}(\pi_{b(i)k}, \boldsymbol{x_i}) = \pi_{b(i)k}a^g_{ik}\Bigl(g_{i0} - r_i^{g,dn} \Bigl)$
        \item $d^*_{jk} = d_{j0} + r_j^{d,dn}$ and $\psi^d_{jk}(\pi_{b(j)k}, \boldsymbol{y_j}) = \pi_{b(j)k}\Bigl(d_{j0} + r_j^{d,dn}\Bigl)$
    \end{itemize}

    \item If $\pi_{bk}$ equals zero, $\psi^g_{ik}(\pi_{b(i)k}, \boldsymbol{x_i})$ and $\psi^d_{jk}(\pi_{b(j)k}, \boldsymbol{y_j})$ will be both equal to zero for the market agents at bus $b$.
\end{enumerate}

Moreover, defining $\pi_{bk}^+$ = $\max\{\pi_{bk}, 0\}$ and $\pi_{bk}^-$ = $-\min\{\pi_{bk}, 0\}$, and using the results presented above, $\psi^g_{ik}(\pi_{b(i)k}, \boldsymbol{x_i})$ and $\psi^d_{jk}(\pi_{b(j)k}, \boldsymbol{y_j})$ can be expressed as:

\vspace{-1.5ex}
\begin{equation}
    \label{eq:psi_generator}
    \psi^g_{ik}(\pi_{b(i)k}, \boldsymbol{x_i}) = \pi_{b(i)k}a^g_{ik}g_{i0} + \pi_{b(i)k}^+a^g_{ik}r^{g,up}_i + \pi_{b(i)k}^-a^g_{ik}r^{g,dn}_i
\end{equation}
\vspace{-2.5ex}
\begin{equation}
    \label{eq:psi_demand}
    \psi^d_{jk}(\pi_{b(j)k}, \boldsymbol{y_j}) = \pi_{b(j)k}d_{j0} - \pi_{b(j)k}^+r^{d,up}_j - \pi_{b(j)k}^-r^{d,dn}_j
\end{equation}

According to the strong duality theorem \citep{Kluwer}, the minimization problem \eqref{eq:EMobj}--\eqref{eq:EMcon13} and the maximization of the Lagrangian dual function \eqref{eq:EM_Lagrangian2} over $\boldsymbol{\Pi}$ yield identical values for their respective objective functions. Additionally, at the optimal dual solution $\boldsymbol{\Pi^*}$, the terms associated with sign-unconstrained variables $\boldsymbol{\theta_0}$ and $\boldsymbol{\theta_k}$, represented by $\phi_{\theta}(\boldsymbol{\Pi}^*)$, are equal to zero \citep{Bertsimas}.

Moreover, note that $\boldsymbol{\pi_0^{f+}}, \boldsymbol{\pi_k^{f+}} \geq \boldsymbol{0}$ and $\boldsymbol{\pi_0^{f-}}, \boldsymbol{\pi_k^{f-}} \leq \boldsymbol{0}$. The complementary slackness condition further ensures that, at the optimal dual solution, entry-wise products $\boldsymbol{\pi_0^{f+*}}\boldsymbol{\pi_0^{f-*}}$ and $\boldsymbol{\pi_k^{f+*}} \boldsymbol{\pi_k^{f-*}}, \forall k \in \mathcal{K}$, are all equal to $\boldsymbol{0}$. Thus, defining $\boldsymbol{\pi_0^{f*}} = \boldsymbol{\pi_0^{f+*}} + \boldsymbol{\pi_0^{f-*}}$ and $\boldsymbol{\pi_k^{f*}} = \boldsymbol{\pi_k^{f+*}} + \boldsymbol{\pi_k^{f-*}}$, the following equalities hold:
\vspace{-1.0ex}
\begin{equation}
    \label{eq:pis0}
    \boldsymbol{\pi_0^{f+*} - \pi_0^{f-*}} = |\boldsymbol{\pi_0^{f*}}|
\end{equation}
\vspace{-2.5ex}
\begin{equation}
    \label{eq:pis}
    \boldsymbol{\pi_k^{f+*} - \pi_k^{f-*}} = |\boldsymbol{\pi_k^{f*}}|, \forall k \in \mathcal{K}
\end{equation}

Consequently, Equation \eqref{eq:EM_Lagrangian4_JM} at the optimal solution is recast as Equation \eqref{eq:EM_Lagrangian5} by using Equations \eqref{eq:psi_generator}--\eqref{eq:pis} and  dropping $\phi_\theta(\boldsymbol{\Pi}^*)$:

\begin{equation}
\begin{aligned}
    \label{eq:EM_Lagrangian5}
    \phi\left(\boldsymbol{\Pi^*}\right) =& - \sum\limits_{l \in \mathcal{L}} \left( \left|\pi_{l0}^{f*}\right| + \sum\limits_{k \in \mathcal{K}} \left|\pi_{lk}^{f*}\right| \right) F_l \\    
    & - \sum\limits_{i \in \mathcal{I}} \max\limits_{\boldsymbol{x_i} \in \mathcal{X}_i} \Bigg[ \left(\pi^*_{b(i)0} + \sum\limits_{k \in \mathcal{K}} \pi^*_{b(i)k} a^g_{ik} - c_i \right) g_{i0} \\
    &+ \left(\sum\limits_{k \in \mathcal{K}}\pi_{b(i)k}^{+*}a^g_{ik} - q_i^{g,up} \right)r^{g,up}_i 
    +\left(\sum\limits_{k \in \mathcal{K}}\pi_{b(i)k}^{-*}a^g_{ik} - q_i^{g,dn} \right)r^{g,dn}_i \Bigg] \\
    &-\sum\limits_{j \in \mathcal{J}}\max\limits_{\boldsymbol{y_j} \in \mathcal{Y}_j}\Bigg[\left(w_j - \pi^*_{b(j)0} - \sum\limits_{k \in \mathcal{K}}\pi^*_{b(j)k} \right)d_{j0}\\
    &+\left(\sum\limits_{k \in \mathcal{K}}\pi_{b(j)k}^{+*} - q_j^{d,up} \right)r^{d,up}_j 
    + \left( \sum\limits_{k \in \mathcal{K}}\pi_{b(j)k}^{-*}  - q_j^{d,dn}\right)r^{d,dn}_j  \Bigg]
\end{aligned}
\end{equation}

Similar to the approach described in Section \ref{sect:First_phase_1}, Equation \eqref{eq:EM_Lagrangian5} can be recast by implicitly considering generation availability statuses $a^g_{ik}$ in set $\mathcal{K}^{OFF}_i$ as follows:

\begin{equation}
\begin{aligned}
    \label{eq:EM_Lagrangian6}
    \phi\left(\boldsymbol{\Pi}^*\right) =& - \sum\limits_{l \in \mathcal{L}} \left( \left|\pi_{l0}^{f*}\right| + \sum\limits_{k \in \mathcal{K}} \left|\pi_{lk}^{f*}\right| \right) F_l \\    
    & - \sum\limits_{i \in \mathcal{I}} \max\limits_{\boldsymbol{x_i} \in \mathcal{X}_i} \Bigg[ \left(\pi^*_{b(i)0} + \sum\limits_{k \in \mathcal{K}} \pi^*_{b(i)k} - c_i \right) g_{i0} \\
    &+ \left(\sum\limits_{k \in \mathcal{K}}\pi_{b(i)k}^{+*} - q_i^{g,up} \right)r^{g,up}_i 
    +\left(\sum\limits_{k \in \mathcal{K}}\pi_{b(i)k}^{-*} - q_i^{g,dn} \right)r^{g,dn}_i  \\
    &-\sum\limits_{k \in \mathcal{K}^{OFF}_{i}} \Bigg( \pi^*_{b{(i)}k}g_{i0} + \pi_{b{(i)}k}^{+*}r_i^{g,up} 
    + \pi_{b{(i)}k}^{-*}r_i^{g,dn} \Bigg) \Bigg] \\  
    &-\sum\limits_{j \in \mathcal{J}}\max\limits_{\boldsymbol{y_j} \in \mathcal{Y}_j}\Bigg[\left(w_j - \pi^*_{b(j)0} - \sum\limits_{k \in \mathcal{K}}\pi^*_{b(j)k}\right)d_{j0}\\
    &+\left( \sum\limits_{k \in \mathcal{K}}\pi_{b(j)k}^{+*}   - q_j^{d,up}\right)r^{d,up}_j 
    + \left( \sum\limits_{k \in \mathcal{K}}\pi_{b(j)k}^{-*} - q_j^{d,dn}   \right)r^{d,dn}_j  \Bigg]
\end{aligned}
\end{equation}

The summation in the first term in the right-hand side of \eqref{eq:EM_Lagrangian6} represents the transmission congestion rent, which is made up of the revenues collected by transmission lines. Note that transmission lines are compensated through the Lagrange multipliers associated with the constraints bounding line power flows. More specifically, transmission prices result from the summation of $\left|\boldsymbol{\pi_0^{f*}}\right|$ and $\sum\limits_{k\in \mathcal{K}}\left|\boldsymbol{\pi_k^{f*}}\right|$.

The total profit of generator $i$ is characterized by the four terms within the first pair of square brackets in \eqref{eq:EM_Lagrangian6}. The first term is the energy profit, from which the energy price at bus $b$ is defined as $\pi^*_{b0}+\sum\limits_{k\in \mathcal{K}}\pi^*_{bk}$, as done in \citet{Arroyo_2005} (Table \ref{tab:EM_pricing_system_arroyo}). Interestingly, the three other generator profit terms involve salient features compared to \citet{Arroyo_2005}. First, from the second and third terms respectively related to up- and down-spinning reserves, separate prices are defined. Thus, up- and down-spinning reserve prices at bus $b$ are equal to $\sum\limits_{k\in \mathcal{K}}\pi_{bk}^{+*}$ and $\sum\limits_{k\in \mathcal{K}}\pi_{bk}^{-*}$, respectively. Additionally, the fourth generator profit term models the security charge. 

Finally, the terms within the second pair of square brackets in \eqref{eq:EM_Lagrangian6} represent the profit of consumer $j$ associated with energy consumption and up- and down-spinning reserve procurement. As can be observed, the above price definitions also hold for the corresponding consumer profit terms.

Tables \ref{tab:price_def1}, \ref{tab:price_def2}, and \ref{tab:price_def3} present the proposed pricing scheme and the corresponding generation and consumer settlements, respectively.

\begin{table}[ht!]
\centering
\caption{Proposed pricing system for problem \eqref{eq:EMobj}--\eqref{eq:EMcon13}}
\label{tab:price_def1}
\scalebox{0.80}{
\begin{tabular}{|l|l|}
\hline
\multirow{2}{*}{\textbf{Nodal Energy Price}}                & \multirow{2}{*}{$p^e_b = \pi^*_{b0} + \sum\limits_{k \in \mathcal{K}}\pi^*_{bk}$}                                                                                     \\
                                                            &                                                                                                                                                                   \\ \hline
\multirow{2}{*}{\textbf{Nodal Up-Spinning Reserve Price}}   & \multirow{2}{*}{$p_b^{up} = \sum\limits_{k \in \mathcal{K}}\pi_{bk}^{+*}$}                                                                                         \\
                                                            &                                                                                                                                                                   \\ \hline
\multirow{2}{*}{\textbf{Nodal Down-Spinning Reserve Price}} & \multirow{2}{*}{$p_b^{dn} = \sum\limits_{k \in \mathcal{K}}\pi_{bk}^{-*}$}                                                                                         \\
                                                            &                                                                                                                                                                   \\ \hline
\multirow{3}{*}{\textbf{Generation Security Charge}}        & \multirow{3}{*}{$C_i^{g,s} = \sum\limits_{k \in \mathcal{K}^{OFF}_{i}}\bigg(\pi^*_{b{(i)}k}g_{i0} + \pi_{b{(i)}k}^{+*}r_i^{g,up} + \pi_{b{(i)}k}^{-*}r_i^{g,dn}\bigg)$} \\
                                                            &                                                                                                                                                                   \\
                                                            &                                                                                                                                                                   \\ \hline
\multirow{2}{*}{\textbf{Transmission Price}}                & \multirow{2}{*}{$p^f_l = \left|\pi_{l0}^{f*}\right| + \sum\limits_{k \in \mathcal{K}}\left|\pi_{lk}^{f*}\right|$}                                                   \\
                                                            &                                                                                                                                                                   \\ \hline
\end{tabular}
}
\end{table}
\FloatBarrier

\begin{table}[ht!]
\centering
\caption{Generation settlement for problem \eqref{eq:EMobj}--\eqref{eq:EMcon13} according to the proposed pricing system}
\label{tab:price_def2}
\scalebox{0.8}{
\begin{tabular}{|l|l|}
\hline
\multirow{2}{*}{\textbf{Generation Energy Revenue}}                                                       & \multirow{2}{*}{$R^{g,e}_i = p^e_{b{(i)}}g_{i0}$}                  \\
                                                                                                          &                                                                \\ \hline
\multirow{2}{*}{\textbf{\begin{tabular}[c]{@{}l@{}}Generation Up-Spinning\\ Reserve Revenue\end{tabular}}}   & \multirow{2}{*}{$R^{g,up}_i = p^{up}_{b(i)}r^{g,up}_{i}$}   \\
                                                                                                          &                                                                \\ \hline
\multirow{2}{*}{\textbf{\begin{tabular}[c]{@{}l@{}}Generation Down-Spinning\\ Reserve Revenue\end{tabular}}} & \multirow{2}{*}{$R^{g,dn}_i = p^{dn}_{b{(i)}}r^{g,dn}_{i}$} \\
                                                                                                          &                                                                \\ \hline
\multirow{2}{*}{\textbf{Generation Total Revenue}}                                                        & \multirow{2}{*}{$R^{g,t}_i = R^{g,e}_i + R^{g,up}_i + R^{g,dn}_i - C^{g,s}_i$} \\
                                                                                                          &                                                                \\ \hline
\multirow{2}{*}{\textbf{Generation Energy Cost}}                                                          & \multirow{2}{*}{$C^{g,e}_i = c_ig_{i0}$}                           \\
                                                                                                          &                                                                \\ \hline
\multirow{2}{*}{\textbf{\begin{tabular}[c]{@{}l@{}}Generation Up-Spinning \\  Reserve Cost\end{tabular}}} & \multirow{2}{*}{$C^{g,up}_i = q_i^{g,up}r_i^{g,up}$}             \\
                                                                                                          &                                                                \\ \hline
\multirow{2}{*}{\textbf{\begin{tabular}[c]{@{}l@{}}Generation Down-Spinning\\ Reserve Cost\end{tabular}}} & \multirow{2}{*}{$C^{g,dn}_i = q_i^{g,dn}r_i^{g,dn}$}             \\
                                                                                                          &                                                                \\ \hline
\multirow{2}{*}{\textbf{Generation Total Cost}}                                                           & \multirow{2}{*}{$C^{g,t}_i = C^{g,e}_i + C^{g,up}_i + C^{g,dn}_i $}        \\
                                                                                                          &                                                                \\ \hline
\multirow{2}{*}{\textbf{Generation Profit}}                                                               & \multirow{2}{*}{$Profit^g_i = R^{g,t}_i - C^{g,t}_i$}                    \\
                                                                                                          &                                                                \\ \hline
\end{tabular}
}
\end{table}
\FloatBarrier

\begin{table}[ht!]
\centering
\caption{Consumer settlement for problem \eqref{eq:EMobj}--\eqref{eq:EMcon13} according to the proposed pricing system}
\label{tab:price_def3}
\scalebox{0.8}{
\begin{tabular}{|l|l|}
\hline
\multirow{2}{*}{\textbf{Consumer Payment for Energy}} & \multirow{2}{*}{$CP_j^e = p^e_{b(j)}d_{j0}$}                 \\
                                                      &                                                              \\ \hline
\multirow{2}{*}{\textbf{\begin{tabular}[c]{@{}l@{}}Consumer Up-Spinning \\ Reserve Revenue\end{tabular}}}  & \multirow{2}{*}{$R^{d,up}_j = p^{up}_{b(j)}r^{d,up}_{j}$}   \\
                                                      &                                                              \\ \hline
\multirow{2}{*}{\textbf{\begin{tabular}[c]{@{}l@{}}Consumer Down-Spinning\\ Reserve Revenue\end{tabular}}} & \multirow{2}{*}{$R^{d,dn}_j = p^{dn}_{b{(j)}}r^{d,dn}_{j}$} \\
                                                      &                                                              \\ \hline
\multirow{2}{*}{\textbf{Consumer Utility}}            & \multirow{2}{*}{$U^d_j = w_jd_{j0}$}                         \\
                                                      &                                                              \\ \hline
\multirow{2}{*}{\textbf{\begin{tabular}[c]{@{}l@{}}Consumer Up-Spinning \\ Reserve Cost\end{tabular}}}     & \multirow{2}{*}{$C^{d,up}_j = q_j^{d,up}r_j^{d,up}$}        \\
                                                      &                                                              \\ \hline
\multirow{2}{*}{\textbf{\begin{tabular}[c]{@{}l@{}}Consumer Down-Spinning\\ Reserve Cost\end{tabular}}}    & \multirow{2}{*}{$C^{d,dn}_j = q_j^{d,dn}r_j^{d,dn}$}        \\
                                                      &                                                              \\ \hline
\multirow{2}{*}{\textbf{Consumer Total Cost}}         & \multirow{2}{*}{$C^{d,t}_j = C^{d,up}_j+C^{d,dn}_j$} \\
                                                      &                                                              \\ \hline
\multirow{2}{*}{\textbf{Consumer Payment}}            & \multirow{2}{*}{$CP_j = CP_j^e - R^{d,up}_j - R^{d,dn}_j$}   \\
                                                      &                                                              \\ \hline
\multirow{2}{*}{\textbf{Consumer Profit}}             & \multirow{2}{*}{$Profit^d_j = U^d_j - C^{d,t}_j - CP_j $}            \\
                                                      &                                                              \\ \hline
\end{tabular}
}
\end{table}
\FloatBarrier

Tables \ref{tab:prices_2bus_Luiza}--\ref{tab:system_settl_2bus_luiza} summarize the results from the application of the proposed pricing scheme to the two-bus example examined in Section \ref{sect: em_example_arroyo}.

\begin{table}[h!]
\caption{Two-bus example -- Proposed nodal energy and reserve prices (\$/MWh)}
\label{tab:prices_2bus_Luiza}
\centering
\resizebox{0.6\textwidth}{!}{
\begin{tabular}{cccc}
\hline
\textbf{$b$} & \textbf{\begin{tabular}[c]{@{}c@{}}Nodal Energy \\ Price, $p_b^e$\end{tabular}} & \textbf{\begin{tabular}[c]{@{}c@{}}Nodal Up-Spinning \\ Reserve Price, $p_b^{up}$\end{tabular}} & \textbf{\begin{tabular}[c]{@{}c@{}}Nodal Down-Spinning \\ Reserve Price, $p_b^{dn}$\end{tabular}} \\ \hline
\textbf{1}   & 200                                                                             & 180                                                                                             & 0                                                                                                 \\
\textbf{2}   & 100                                                                             & \textcolor{white}{0}85                                                                          & 5                                                                                                 \\ \hline
\end{tabular}
}
\end{table}
\begin{table}[h!]
\caption{Two-bus example -- Values of $\boldsymbol{\pi_0^{f*}}$, $\boldsymbol{\pi_k^{f*}}$, and transmission price (\$/MWh)}
\label{tab:lagrange_line_2bus}
\renewcommand{\arraystretch}{1.2} 
\centering
\resizebox{0.7\textwidth}{!}{
\begin{tabular}{ccccccc}
\hline
\multicolumn{1}{l}{\textbf{}} & \multicolumn{1}{l}{}            & \multicolumn{4}{c}{$\boldsymbol{\pi_k^{f*}}$}                                                                                                                                                                                                                                                                  & \multicolumn{1}{l}{}                                                           \\ \cline{3-6}
\textbf{$l$}                 & \textbf{$\boldsymbol{\pi_0^{f*}}$} & \textbf{\begin{tabular}[c]{@{}c@{}}Outage of \\ Generator 1\end{tabular}} & \textbf{\begin{tabular}[c]{@{}c@{}}Outage of \\ Generator 2\end{tabular}} & \textbf{\begin{tabular}[c]{@{}c@{}}Outage of \\ Generator 3\end{tabular}} & \textbf{\begin{tabular}[c]{@{}c@{}}Outage of \\ Line 1--2\end{tabular}} & \textbf{\begin{tabular}[c]{@{}c@{}}Transmission Price,\\ $p^f_l$\end{tabular}} \\ \hline
\textbf{1--2}                 & 0                               & 95                                                                        & 0                                                                         & 0                                                                         & -                                                                       & 95                                                                             \\ \hline
\end{tabular}
}

\end{table}
\begin{table}[h!]
\captionsetup{justification=centering, width=\linewidth}
\caption{Two-bus example -- Proposed generation settlement (\$)}
\label{tab:revenues_gen_2bus_Luiza}
\centering
\resizebox{0.8\textwidth}{!}{
\renewcommand{\arraystretch}{1.2} 
\begin{tabular}{ccccccccccc}
\hline
$i$ &
  $R^{g,e}_i$ &
  $R_i^{g,up}$ &
  $R_i^{g,dn}$ &
  $C^{g,s}_i$ &
  $R^{g,t}_i$ &
  $C^{g,e}_i$ &
  $C^{g,up}_i$ &
  $C^{g,dn}_i$ &
  $C^{g,t}_i$ &
  $Profit^g_i$ \\ \hline
\textbf{1} &
  15,000 &
  \textcolor{white}{000,}0 &
  \textcolor{white}{0}0 &
  13,500 &
  \textcolor{white}{0}1,500 &
  1,500 &
  \textcolor{white}{00}0 &
  \textcolor{white}{0}0 &
  1,500 &
  \textcolor{white}{000,}0 \\
\textbf{2} &
  \textcolor{white}{0}3,000 &
  2,550 &
  25 &
  \textcolor{white}{0000,}0 &
  \textcolor{white}{0}5,575 &
  1,500 &
  150 &
  25 &
  1,675 &
  3,900 \\
\textbf{3} &
  \textcolor{white}{0}1,500 &
  2,975 &
  \textcolor{white}{0}0 &
  \textcolor{white}{0000,}0 &
  \textcolor{white}{0}4,475 &
  1,500 &
  350 &
  \textcolor{white}{0}0 &
  1,850 &
  2,625 \\ \hline
\textbf{Total} &
  \textbf{19,500} &
  \textbf{5,525} &
  \textbf{25} &
  \textbf{13,500} &
  \textbf{11,550} &
  \textbf{4,500} &
  \textbf{500} &
  \textbf{25} &
  \textbf{5,025} &
  \textbf{6,525} \\ \hline
\end{tabular}
}
\end{table}
\begin{table}[h!]
\captionsetup{justification=centering, width=\linewidth}
\caption{Two-bus example -- Proposed consumer settlement (\$)}
\label{tab:revenues_consumers_2bus_Luiza}
\centering
\resizebox{0.8\textwidth}{!}{
\renewcommand{\arraystretch}{1.2} 
\begin{tabular}{cccccccccc}
\hline
$j$        & $CP^e_j$ & $R^{d,up}_j$ & $R^{d,dn}_j$ & $CP_j$ & $U^d_{j}$ & $C^{d,up}_j$ & $C^{d,dn}_j$ & $C^{d,t}_j$               & $Profit^d_j$             \\ \hline
\textbf{1} & 16,000   & 1,800        & 0            & 14,200 & 16,000    & 1,500        & 0            & \textcolor{white}{0}1,500 & \textcolor{white}{0,}300 \\
\textbf{2} &
  \textcolor{white}{0}4,000 &
  \textcolor{white}{000,}0 &
  0 &
  \textcolor{white}{0}4,000 &
  \textcolor{white}{0}6,000 &
  \textcolor{white}{000,}0 &
  0 &
  \textcolor{white}{0000,}0 &
  2,000 \\ \hline
\textbf{Total} &
  \textbf{20,000} &
  \textbf{1,800} &
  \textbf{0} &
  \textbf{18,200} &
  \textbf{22,000} &
  \textbf{1,500} &
  \textbf{0} &
  \textcolor{white}{0}\textbf{1,500} &
  \textbf{2,300} \\ \hline
\end{tabular}
}
\end{table}
\begin{table}[h!]
\captionsetup{justification=centering, width=\linewidth}
\caption{Two-bus example -- Proposed system settlement (\$)}
\label{tab:system_settl_2bus_luiza}
\centering
\resizebox{0.3\textwidth}{!}{
\begin{tabular}{cc}
\hline
\textbf{Generation Revenue} & 11,550                    \\
\textbf{Transmission Revenue}       & \textcolor{white}{0}6,650 \\
\textbf{Consumer Payment}   & 18,200                    \\ \hline
\textbf{Balance}            & \textcolor{white}{0000} 0 \\ \hline
\end{tabular}
}
\end{table}

As can be seen in Table \ref{tab:prices_2bus_Luiza}, the proposed nodal energy prices are the same as those attained by the methodology presented in \citet{Arroyo_2005}, thereby leading to identical energy revenues for generators and energy payments by consumers. By contrast, the use of different nodal prices for up- and down-spinning reserves (Table \ref{tab:prices_2bus_Luiza}) constitutes a significant departure from the single nodal price defined in \citet{Arroyo_2005} for both services, thus yielding substantial differences for the corresponding revenues. For this particular example, the comparison of Tables \ref{tab:revenues_gen_2bus_Luiza} and \ref{tab:revenues_gen_2bus_Arroyo} shows that generators 2 and 3 increase their up-spinning reserve revenues by $6.25 \%$ whereas the down-spinning reserve revenue of generator 2 decreases by a factor of 16. Additionally, as generator 1 is responsible for the need for up-spinning reserve, a security charge is levied on this generator.

As for consumers, the settlement remains unaltered for this particular example (Tables \ref{tab:revenues_consumers_2bus_Luiza} and \ref{tab:revenues_consumers_2bus_Arroyo}). Note that the only load contributing to security is load 1 at bus 1, in the form of up-spinning reserve. For this bus, the proposed up-spinning reserve price happens to be the same as the security price resulting from the method described in \citet{Arroyo_2005}, thereby giving rise to identical revenues.  

Moreover, unlike \citet{Arroyo_2005}, line congestion under contingency is acknowledged by the proposed transmission price $p^f_l$ (Table \ref{tab:lagrange_line_2bus}) and the corresponding transmission revenue, $p^f_lF_l$, which, for this case, amounts to $95 \times 70 = \$6,650$.

As expected, the total generation profit, the total consumer profit, and the transmission revenue sum up the aforementioned optimal social welfare, i.e., $\$15,475$. Remarkably, unlike the pricing scheme described in \citet{Arroyo_2005}, the proposed method renders the total consumer payment equal to the sum of the generation revenue and the transmission revenue (Table \ref{tab:system_settl_2bus_luiza}), as is desirable. These results provide empirical support for the revenue adequacy and revenue neutrality featured by the proposed pricing scheme for the market-clearing problem analyzed in this section. The proofs for such extensions of Theorems 1 and 2 are provided in Appendices D and E, respectively.

\section{Conclusion}
\label{sect:conclusion}

This paper has presented a new causation-based pricing framework as an alternative to a previously reported scheme. Major salient features include explicitly defining prices for up-spinning reserves, down-spinning reserves, and transmission services, along with a novel security charge mechanism. These additions ensure a more comprehensive allocation of the costs associated with operating reserves and system reliability, resulting in efficient market operations. The proposed approach is rigorously grounded in the Lagrangian dual function, leveraging Lagrangian multipliers to establish nodal prices and security charges. More importantly, the proposed pricing scheme yields a market settlement featuring two relevant and desirable properties, namely revenue adequacy and revenue neutrality, thereby avoiding the need for \textit{ad hoc} out-of-market adjustments.

Numerical results corroborate the findings in the related literature, de\-monstrating that larger generators tend to bear higher security charges due to their significant contribution to the system-wide reserve requirements under contingency. Interestingly, these charges keep the same incentives of the uniform pricing system, largely used for energy and reserves, while ensuring specific reliability incentives based on the cost-causation principle. Additionally, the newly proposed pricing framework provides relevant incentives for transmission assets, remunerating them based on energy and reserve utilization across all contingency states. This unification guarantees a consistent and transparent cost allocation framework.

This work paves the way for future research in this field. One potential direction is the extension of the proposed model to a multi-period setting, allowing for the explicit consideration of inter-temporal aspects in market operations. Additionally, the incorporation of renewable generation uncertainty would allow the pricing framework to better reflect the variability and reliability challenges introduced by increasing renewables penetration. Finally, further investigation into economic incentives could provide deeper insights into how market participants respond to security charges and reserve pricing.

\appendix

\section{Nomenclature}
\vspace{-5ex}
\renewcommand{\nomname}{} 
\input{main.nls}
\section{Proof of Revenue Adequacy for the Pricing Scheme of Section 2.3}
\label{sect:proof1}

Let us rewrite the LD function in Equation \eqref{eq:SM_Lagrangian5} as follows:
\begin{equation}
    \begin{aligned}
        \label{eq:SM_Lagrangian6_app}
        \phi(\boldsymbol{\Pi}) = & \ CP(\boldsymbol{\Pi}) 
        - \sum\limits_{i \in \mathcal{I}} \max\limits_{\boldsymbol{x}_i \in \mathcal{X}_i} \Bigg\{ Profit^g_i(\boldsymbol{\Pi}, \boldsymbol{x_i}) \Bigg\}
    \end{aligned}
\end{equation}

\noindent
where $CP(\boldsymbol{\Pi})$ and $Profit^g_i(\boldsymbol{\Pi}, \boldsymbol{x_i})$ represent the consumer payment and the profit of generator $i$, respectively. Since $Profit^g_i(\boldsymbol{\Pi},\boldsymbol{x_i})$ is 
a linear objective function maximized over $\boldsymbol{x_i}$, with $\boldsymbol{x_i} \geq \boldsymbol{0}$, the optimal value of $Profit^g_i(\boldsymbol{\Pi}, \boldsymbol{x_i})$ is guaranteed to be non-negative. In other words, if $\pi_0$ and $\pi_k$ fail to cover generator $i$'s costs, the generator will not produce any output, resulting in $g^*_{i0} = r^{g,up*}_i = 0$. Likewise, at the optimal solution $\boldsymbol{x^*_i}$ and $\boldsymbol{\Pi^*}$:

\begin{equation}
    Profit^g_i(\boldsymbol{\Pi^*},\boldsymbol{x^*_i}) \geq 0, \ \forall i \in \mathcal{I}
\end{equation}

Therefore, the theorem holds.
\hfill $\qed$

\section{Proof of Revenue Neutrality for the Pricing Scheme of Section 2.3}
\label{sect:proof2}
By the strong duality theorem, the optimal value of the primal objective function in Equation \eqref{eq:SMobj} is equal to the optimal value of the LD function in Equation \eqref{eq:SM_Lagrangian6}. Thus, at the optimal solution $\boldsymbol{x}^*$ and $\boldsymbol{\Pi}^*$, we have:

\begin{equation}
\begin{aligned}
    \label{sect:app-LDP}
    \sum\limits_{i \in \mathcal{I}}c_ig_{i0}^* + \sum\limits_{i \in \mathcal{I}}q^{g,up}_ir_i^{g,up*} &= \left(\pi^*_0 + \sum\limits_{k \in \mathcal{K}} \pi^*_k \right) d \\
    &\quad - \sum\limits_{i \in \mathcal{I}} \Bigg[ 
        \left(\pi^*_{0} + \sum\limits_{k \in \mathcal{K}} \pi^*_k - c_i \right) g^*_{i0} \\
    &\quad + \left(\sum\limits_{k \in \mathcal{K}} \pi^*_k - q_i^{g,up} \right) r_i^{g,up*} \\
    &\quad - \sum\limits_{k \in \mathcal{K}^{OFF}_i} \pi^*_k \left(g_{i0}^* + r_i^{g,up*}\right) \Bigg]
\end{aligned}
\end{equation}

By canceling out identical terms in both sides of Equation \eqref{sect:app-LDP} and rearranging terms, we simplify the expression as follows:

\begin{equation}
\begin{aligned}
    \label{sect:app-LDP2}
    \left(\pi^*_0 + \sum\limits_{k \in \mathcal{K}} \pi^*_k \right) d &= 
     \sum\limits_{i \in \mathcal{I}} \Bigg[ 
        \left(\pi^*_{0} + \sum\limits_{k \in \mathcal{K}} \pi^*_k \right) g^*_{i0} \\
    &\quad + \sum\limits_{k \in \mathcal{K}} \pi^*_k  r_i^{g,up*} 
    - \sum\limits_{k \in \mathcal{K}^{OFF}_i} \pi^*_k \left(g_{i0}^* + r_i^{g,up*}\right) \Bigg]
\end{aligned}
\end{equation}

Therefore, at the optimal solution, the consumer payment equals the sum of all generator revenues.
\qed
\section{Proof of Revenue Adequacy for the Pricing Scheme of Section 3.3}
\label{sect:proof3}

Let us rewrite the LD function in Equation \eqref{eq:EM_Lagrangian6} as follows:

\begin{equation}
\begin{aligned}
    \label{eq:EM_Lagrangian5_app}
    \phi\left(\boldsymbol{\Pi}^*\right) =& - \sum\limits_{l \in \mathcal{L}} \left( \left|\pi_{l0}^{f*}\right| + \sum\limits_{k \in \mathcal{K}} \left|\pi_{lk}^{f*}\right| \right) F_l \\    
    & - \sum\limits_{i \in \mathcal{I}} \max\limits_{\boldsymbol{x_i} \in \mathcal{X}_i} \Bigg[ Profit^g_i\left(\boldsymbol{\Pi^*_{b(i)}}, \boldsymbol{x_i}\right) \Bigg] \\
    &-\sum\limits_{j \in \mathcal{J}}\max\limits_{\boldsymbol{y_j} \in \mathcal{Y}_j}\Bigg[Profit^d_j\left(\boldsymbol{\Pi^*_{b(j)}}, \boldsymbol{y_j}\right)  \Bigg]
\end{aligned}
\end{equation}

The terms $Profit^g_i\left(\boldsymbol{\Pi_{b(i)}^*}, \boldsymbol{x_i}\right)$ and $Profit^d_j\left(\boldsymbol{\Pi_{b(j)}^*}, \boldsymbol{y_j}\right)$ respectively represent the profits of generator $i$ and consumer $j$. It should be noted that $Profit^g_i\left(\boldsymbol{\Pi_{b(i)}^*}, \boldsymbol{x_i}\right)$ and $Profit^d_j\left(\boldsymbol{\Pi_{b(j)}^*}, \boldsymbol{y_j} \right)$ are linear objective functions respectively maximized over $\boldsymbol{x_i}$ and $\boldsymbol{y_j}$ with $\boldsymbol{x_i} \geq \boldsymbol{0}$ and $\boldsymbol{y_j} \geq \boldsymbol{0}$. Therefore, at the optimal solution, $Profit^g_i\left(\boldsymbol{\Pi_{b(i)}^*}, \boldsymbol{x^*_i}\right)$ and $Profit^d_j\left(\boldsymbol{\Pi_{b(j)}^*}, \boldsymbol{y^*_j}\right)$ are both non-negative. In other words, if $\pi^*_{b(i)0}$ and $\pi^*_{b(i)k}$ fail to cover generator $i$'s costs, the production and reserve contributions of this generator will be $0$, i.e., $g^*_{i0} = r^{g,up*}_i = r^{g,dn*}_i = 0$. Analogously, if $\pi^*_{b(j)0}$ and $\pi^*_{b(j)k}$ are not profitable for consumer $j$, the consumption and reserve contributions of this consumer will be $0$, i.e., $d^*_{j0} = r^{d,up*}_j = r^{d,dn*}_j = 0$. Therefore:

\begin{equation}
    Profit^g_i\left(\boldsymbol{\Pi_{b(i)}^*}, \boldsymbol{x^*_i}\right) \geq 0, \ \forall i \in \mathcal{I} 
\end{equation}
\vspace{-3ex}
\begin{equation}
    Profit^d_j\left(\boldsymbol{\Pi_{b(j)}^*}, \boldsymbol{y^*_j}\right)  \geq 0, \ \forall j \in \mathcal{J}
\end{equation}

Therefore, the theorem holds.
\hfill $\qed$

\section{Proof of Revenue Neutrality for the Pricing Scheme of Section 3.3}
\label{sect:proof4}

By the strong duality theorem, the optimal value of the primal objective function in Equation \eqref{eq:EMobj} is equal to the optimal value of the LD function in Equation \eqref{eq:EM_Lagrangian6}. Thus, at the optimal solution $\boldsymbol{x^*}$, $\boldsymbol{y^*}$, and $\boldsymbol{\Pi^*}$, we have:

\begin{equation}
\label{eq:app_lulu}
\begin{aligned}
    & \sum_{i \in \mathcal{I}} c_i g^*_{i0} 
    + \sum_{i \in \mathcal{I}} q_i^{g,up} r_i^{g,up*} 
    + \sum_{i \in \mathcal{I}} q_i^{g,dn} r_i^{g,dn*} \\
    &
    - \sum_{j \in \mathcal{J}} w_j d^*_{j0}
    + \sum_{j \in \mathcal{J}} q_j^{d,up} r_j^{d,up*} 
    + \sum_{j \in \mathcal{J}} q_j^{d,dn} r_j^{d,dn*} = \\
    & 
    -\sum_{l \in \mathcal{L}} \left( |\pi_{l0}^{f*}| + \sum_{k \in \mathcal{K}} |\pi_{lk}^{f*}| \right) F_l     
    -\sum_{i \in \mathcal{I}} 
    \Bigg[ 
        \left( 
            \pi^*_{b(i)0} + \sum_{k \in \mathcal{K}} \pi^*_{b(i)k} - c_i 
        \right) g^*_{i0} \\
    & \quad 
        + \left( 
            \sum_{k \in \mathcal{K}} \pi_{b(i)k}^{+*} - q_i^{g,up} 
        \right) r_i^{g,up*} 
        + \left( 
            \sum_{k \in \mathcal{K}} \pi_{b(i)k}^{-*} - q_i^{g,dn} 
        \right) r_i^{g,dn*} \\
    & \quad 
        - \sum_{k \in \mathcal{K}_{i}^{\text{OFF}}} 
        \Bigg( 
            \pi^*_{b(i)k} g^*_{i0} 
            + \pi_{b(i)k}^{+*} r_i^{g,up*} 
            + \pi_{b(i)k}^{-*} r_i^{g,dn*} 
        \Bigg) 
    \Bigg] \\
    &- \sum_{j \in \mathcal{J}} 
    \Bigg[ 
        \left( 
            w_j - \pi^*_{b(j)0} - \sum_{k \in \mathcal{K}} \pi^*_{b(j)k} 
        \right) d^*_{j0} \\
    & \quad 
        + \left( 
            \sum_{k \in \mathcal{K}} \pi_{b(j)k}^{+*} - q_j^{d,up} 
        \right) r_j^{d,up*} 
        + \left( 
            \sum_{k \in \mathcal{K}} \pi_{b(j)k}^{-*} - q_j^{d,dn} 
        \right) r_j^{d,dn*} 
    \Bigg]
\end{aligned}
\end{equation}

By canceling out identical terms in both sides of Equation \eqref{eq:app_lulu} and rearranging terms, we simplify the expression as follows:

\allowdisplaybreaks

\begin{equation}
\begin{aligned}
    &\sum_{j \in \mathcal{J}} 
        \Bigg[ 
            \left( 
                \pi^*_{b(j)0} + \sum_{k \in \mathcal{K}} \pi^*_{b(j)k} 
            \right) d^*_{j0}  
            - \sum_{k \in \mathcal{K}} \pi_{b(j)k}^{+*} r_j^{d,up*} 
            - \sum_{k \in \mathcal{K}} \pi_{b(j)k}^{-*} r_j^{d,dn*} 
        \Bigg] = \\
    &\sum_{l \in \mathcal{L}} \left( |\pi_{l0}^{f*}| + \sum_{k \in \mathcal{K}} |\pi_{lk}^{f*}| \right) F_l 
    + \sum_{i \in \mathcal{I}} \Bigg[ 
        \left( 
            \pi^*_{b(i)0} + \sum_{k \in \mathcal{K}} \pi^*_{b(i)k} 
        \right) g^*_{i0} 
        + \sum_{k \in \mathcal{K}} \pi_{b(i)k}^{+*} r_i^{g,up*} \\
        & \quad
        + \sum_{k \in \mathcal{K}} \pi_{b(i)k}^{-*} r_i^{g,dn*}  
        - \sum_{k \in \mathcal{K}_{i}^{\text{OFF}}} 
        \Bigg( 
            \pi^*_{b(i)k} g^*_{i0} 
            + \pi_{b(i)k}^{+*} r_i^{g,up*} 
            + \pi_{b(i)k}^{-*} r_i^{g,dn*} 
        \Bigg) 
    \Bigg] 
\end{aligned}
\end{equation}

Therefore, at the optimal solution, the sum of all consumer payments equals the sum of all generator and transmission line revenues.
\hfill $\qed$

\section*{Acknowledgement}
The work of Alexandre Street and Luíza Ribeiro is partly funded by the World Bank's Project BR-CCEE-TDR-14-21-PRECO-CS-QBS, within the scope of the META II initiative and with the support of the Brazilian Electricity Trading Chamber (CCEE) and PSR.

\bibliographystyle{apalike}

\bibliography{References}

\end{document}